\documentclass{article}

\usepackage{arxiv}

\usepackage[utf8]{inputenc} 
\usepackage[T1]{fontenc}    
\usepackage{hyperref}       
\usepackage{url}            
\usepackage{booktabs}       
\usepackage{amsfonts}       
\usepackage{nicefrac}       
\usepackage{microtype}      
\usepackage{lipsum}		
\usepackage{graphicx}
\usepackage[numbers]{natbib}
\usepackage{doi}

\usepackage{enumitem}
\usepackage{tabularx} 
\usepackage{multirow}
\usepackage{caption}
\usepackage{subcaption}
\usepackage{float}
\usepackage{siunitx}

\usepackage{chngcntr} 
\usepackage{tocloft} 

\usepackage{array}
\newcolumntype{Y}{>{\raggedright\arraybackslash}X}

\DeclareSIUnit\years{years}

\usepackage{xcolor}
\newcommand{\hrulegray}{%
  \vspace{1.0ex}
}

\providecommand{\Description}[1]{}

\title{Is Seeing Believing? Evaluating Human Sensitivity to Synthetic Video}

\date{March, 2026}

\author{ \href{https://orcid.org/0000-0002-7372-9850}{\includegraphics[scale=0.06]{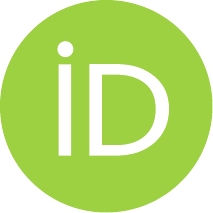}\hspace{1mm}David Wegmann}\\
	Department of Media and Journalism Studies \\
	School of Communication and Culture\\
    Aarhus University \\
	Aarhus, Denmark \\
	\texttt{david.wegmann@cc.au.dk} \\
	   \And
	Emil Stevnsborg \\
	Department of Computer Science\\
	University of Copenhagen\\
	Copenhagen, Denmark \\
	\texttt{emilstevnsborg.official@gmail.com} \\
		\And
    \href{https://orcid.org/0000-0002-8306-1102}{\includegraphics[scale=0.06]{orcid.pdf}\hspace{1mm}Søren Knudsen}\\
	IT University of Copenhagen\\
	Copenhagen, Denmark \\
	\texttt{soekn@itu.dk} \\
    	\And
	\href{https://orcid.org/0000-0002-3629-2039}{\includegraphics[scale=0.06]{orcid.pdf}\hspace{1mm}Luca Rossi}\\
	IT University of Copenhagen\\
	Copenhagen, Denmark \\
	\texttt{lucr@itu.dk} \\
    	\And
	\href{https://orcid.org/0000-0003-1827-8513}{\includegraphics[scale=0.06]{orcid.pdf}\hspace{1mm}Aske Mottelson}\\
	IT University of Copenhagen\\
	Copenhagen, Denmark \\
	\texttt{asmo@itu.dk} \\
}

\renewcommand{\headeright}{}
\renewcommand{\undertitle}{}
\renewcommand{\shorttitle}{Is Seeing Believing? Evaluating Human Sensitivity to Synthetic Video}

\hypersetup{
pdftitle={Is Seeing Believing? Evaluating Human Sensitivity to Synthetic Video},
pdfsubject={cs.HC},
pdfauthor={David Wegmann, Emil Stevnsborg, Søren Knudsen, Luca Rossi, Aske Mottelson},
pdfkeywords={Deepfake, misinformation, social media, synthetic media, disfluency, cognitive psychology, experimental psychology},
}

\begin{document}
\maketitle

\begin{abstract}
Advances in machine learning have enabled the creation of realistic synthetic videos known as \emph{deepfakes}. As deepfakes proliferate, concerns about rapid spread of disinformation and manipulation of public perception are mounting. Despite the alarming implications, our understanding of how individuals perceive synthetic media remains limited, obstructing the development of effective mitigation strategies. 
This paper aims to narrow this gap by investigating human responses to visual and auditory distortions of videos and deepfake-generated visuals and narration. In two between-subjects experiments, we study whether audio–visual distortions affect cognitive processing, such as subjective credibility assessment and objective learning outcomes. A third study reveals that artifacts from deepfakes influence credibility. The three studies show that video distortions and deepfake artifacts can reduce credibility. Our research contributes to the ongoing exploration of the cognitive processes involved in the evaluation and perception of synthetic videos, and underscores the need for further theory development concerning deepfake exposure.
\end{abstract}


\section{Introduction}
Recent advances in machine learning architectures capable of synthesizing highly realistic video and audio material, so-called \emph{deepfakes}, threaten the integrity of videos as reliable information sources~\cite{westerlund_emergence_2019}. A central concern is that malicious actors can use the technology to create and spread false evidence, as deepfakes enable the creation of convincing impersonations of influential and trusted figures~\cite{kobis_fooled_2021, Mink2024}, and are frequently deployed in contexts of sexual harassment~\cite{Wolfe2023, Gebru2024}. The introduction of deepfake videos raises concerns as people seem to be particularly susceptible to video-based disinformation~\cite{sundar_seeing_2021, lee_something_2022}. Despite the growing awareness of issues concerning deepfakes, we lack a general understanding of how people perceive and evaluate synthetically generated media~\cite{Tahir2021, Gamage2022}.

The visual aspect of a video may influence how its message is perceived in several ways: It may contain arguments, focus or distract the audience's attention, and serve as cue for heuristic evaluations~\cite{peng_agenda_2023}. A key heuristic in processing visual information is \textit{processing fluency}, the subjective ease of processing a stimulus. When information is easier (or harder) to process than expected, this metacognitive ``feel'' systematically shifts judgments such as truth, trust, and liking~\cite{whittlesea_source_2000}.
Prior research has found evidence for the influence of processing fluency on judgments of credibility and trust in numerous media types~\cite{schwarz_metacognitive_2021}. The legibility of text~\cite{reber_effects_1999}, the quality of audio recordings~\cite{newman_good_2018}, and the presence of nonprobative photos alongside a statement~\cite{newman_nonprobative_2012} were found to affect how credible people deemed information to be. Nevertheless, the effects of processing fluency of videos remain mostly underexplored. 

Given the unique characteristics of videos and the increasing prevalence of deepfakes, studying the effects of visual and auditory video aspects on credibility becomes crucial~\cite{Gamage2022}. This is because artifacts remain the major indicator of artificially generated content, despite the increasing quality of deepfake videos~\cite{ciftci_fakecatcher_2020}. 
These artifacts often manifest as unnatural movements, extraneous limbs, inconsistencies in the subject's lighting, and similar features that do not occur in recorded videos~\cite{matern_exploiting_2019}. 

Distortions and deepfake artifacts can reduce processing fluency because they violate viewers' expectations about how recorded videos typically look and sound. Such low-level irregularities act as perceptual expectancy violations that make the video subjectively harder to process, and may therefore serve as heuristic cues in credibility judgments. This is especially relevant in online video environments, where high information density and limited attention promote intuitive, heuristic modes of evaluation~\cite{greifeneder_when_2011,lang_limited_2000}. Under these conditions, which are common on platforms such as YouTube and TikTok~\cite{metzger_social_2010,sundar_seeing_2021}, viewers are likely to rely on directly accessible perceptual cues rather than analytical assessment. We therefore examine whether visual distortions, auditory distortions, and deepfake artifacts in faces and voices systematically affect video credibility, processing fluency, and related evaluation outcomes.

For HCI, credibility and authenticity are interactional judgments shaped by interface cues as much as by content. Online videos are typically encountered in environments that impose cognitive load and encourage heuristic processing~\cite{lang_limited_2000, metzger_psychological_2015}, meaning that subtle audiovisual irregularities become salient inputs to users' credibility assessments. Yet, we lack empirical evidence on how users in realistic viewing conditions respond to artifacts in synthetic media~\cite{hilligoss_developing_2008, sundar_main_2008, metzger_psychological_2015}.

In this paper, we address three questions: (1) do visual distortions reduce video credibility; (2) do auditory distortions and audio–visual timing irregularities have similar effects; and (3) do deepfake artifacts in faces and voices, separately and combined, undermine credibility and learning? Across three preregistered online experiments we find that visual distortions and deepfake artifacts reliably lower message credibility and source vividness, while leaving learning largely unaffected.

We find that deepfake artifacts reduce both message credibility and perceived source vividness, while processing fluency remains unaffected. Our contributions are: (1) empirical evidence that deepfake artifacts undermine credibility in subtle but measurable ways; (2) an analysis of how distortions and deepfake artifacts in audio and video shape user judgments; and (3) open and reproducible research resources, including all data, analyses, and stimuli shared on OSF \url{https://doi.org/10.17605/OSF.IO/T3AWB}). Together, these findings advance our understanding of human sensitivity to deepfakes and provide a basis for developing theory, policy, and interventions for deepfake exposure.o

\section{Background and related work}
\label{sec:background}

\subsection{Deepfakes and Artifacts}
Deepfakes are videos in which faces and/or voices have been synthetically generated using deep-learning models. The technology enables convincing impersonations with comparatively little expertise~\cite{Kaate2023,roose_here_2018} and is increasingly used in deceptive contexts~\cite{dobber_microtargeted_2021}. For the present work we distinguish three categories: (1) \textit{recorded videos}, captured without AI synthesis; (2) \textit{distorted videos}, where visual or auditory degradations are added without using generative models; and (3) \textit{deepfake videos}, in which the facial appearance and/or narration have been synthesized.

Deepfake generation pipelines introduce characteristic artifacts because models must estimate facial geometry, motion, illumination, and texture, and temporally align synthesized audio and video. Imperfections in these steps produce residual cues that differentiate synthetic from recorded footage, such as subtle inconsistencies in shading, boundary alignment, facial dynamics, or mouth–speech synchrony~\cite{matern_exploiting_2019,li_face_2020,yang_exposing_2019,ciftci_fakecatcher_2020}. While some artifacts may be imperceptible to viewers but detectable algorithmically~\cite{agarwal_detecting_2020}, others manifest as visible irregularities or atypical movement patterns.

Despite these cues, viewers often struggle to reliably distinguish deepfakes from recorded videos. Prior studies show that many people fail to identify deepfakes even when aware of their existence~\cite{korshunov_subjective_2021,groh_deepfake_2022,kobis_fooled_2021,Mink2024}, and deepfake political videos frequently receive moderate credibility ratings~\cite{lee_something_2022,hwang_effects_2021}. Uncertainty is widespread, and judgments are easily confounded by factors such as familiarity, liking, or sympathy for the depicted individual~\cite{vaccari_deepfakes_2020,kobis_fooled_2021}. This difficulty underscores the need to understand how specific kinds of visual and auditory artifacts shape users' evaluations.

\subsection{Heuristics and processing fluency in credibility judgments}
Credibility refers to an individual's judgment of the veracity of the content of communication~\cite{appelman_measuring_2016}. Online environments complicate such judgments: information volume is high, source cues are sparse or ambiguous, and professional verification is limited~\cite{callister_media_2000,metzger_making_2007,metzger_psychological_2015,Sohrawardi2024}. Under these conditions of overload and limited cognitive resources, people rely on heuristics; that is, readily accessible cues that can be evaluated quickly but are fallible~\cite{lang_limited_2000,hilligoss_developing_2008}. Social indicators such as endorsements, comments, or source reputation often serve this function~\cite{metzger_social_2010}, and popularity has been shown to be influential for credibility assessments explicitly in the context of deepfake content~\cite{jin_assessing_2023}.

Deepfakes complicate credibility judgments further because they can appropriate trusted identities and obscure more diagnostic cues. Although synthetic videos may appear plausible, they often contain subtle irregularities. Such unexpected visual or auditory features create \emph{expectancy violations}; deviations from what viewers assume recorded videos should look or sound like, which can reduce \emph{processing fluency}, the subjective ease of processing a stimulus~\cite{whittlesea_source_2000}. Processing fluency reliably influences credibility judgments across media: less legible text~\cite{reber_effects_1999} and lower-quality audio~\cite{newman_good_2018} reduce perceived truthfulness~\cite{schwarz_metacognitive_2021}.

People particularly rely on fluency when external knowledge is limited~\cite{alter_uniting_2009}. One proposed mechanism is familiarity: fluently processed information feels more familiar and is therefore judged as more likely to be true~\cite{dechene_truth_2010,whittlesea_illusions_1990}. In the context of deepfakes, this suggests that viewers may turn to the ``feel'' of ease or difficulty when evaluating unfamiliar videos or presenters. However, it remains unclear whether and how such fluency-based heuristics operate for \emph{videos}, and especially for deepfake-generated faces and voices, where multiple perceptual channels may produce competing or ambiguous cues.

\subsection{Credibility assessment as an interaction problem}
In online environments, credibility judgments are shaped by an interactional process that includes the viewer and the design of digital interfaces. Users evaluate authenticity within feeds, video players, and recommendation systems that structure attention and impose cognitive constraints~\cite{hilligoss_developing_2008, metzger_psychological_2015}. HCI research has shown that interface-level cues such as modality, visual clarity, and system agency systematically influence heuristic credibility judgments~\cite{sundar_main_2008,sun2024trust,zieglmeier2021designing}. In this perspective, deepfake artifacts can be understood as perceptual cues embedded in the interaction environment: they are surfaced to the user through the interface in the same way as other credibility-relevant signals. Studying how users respond to these artifacts therefore addresses the central question of how interactive systems mediate trust and uncertainty when synthetic media are encountered in real-world interface contexts.

\section{Studies I--III}
Across three preregistered online experiments, we systematically examined how manipulations to visual and auditory properties of an educational video, including deepfake artifacts, affect processing fluency, credibility, and learning, in addition to select metacognitive outcomes. Study~I tested the hypothesis that reducing the processing fluency of video imagery makes it less credible. Study~II extended this approach to speech clarity and audio--visual synchrony, and added learning and additional metacognitive outcomes. Studies~I and II showed that visual distortions affect perceived processing fluency and credibility. Study~III extended this progression to ecologically realistic deepfake artifacts by independently manipulating synthesized faces and synthesized narration. 

\section{Methods}

\subsection{Design}

All three studies employed between-subjects experimental designs with a single independent variable. Participants were randomly assigned to one condition that differed in the presentation style of an educational video.\\[.15cm]
\noindent
\textbf{Study~I} used a two-level design, comparing a \textsc{baseline} condition with a \textsc{reduced visual clarity} condition. Study~I tested the hypothesis:
\begin{enumerate}[label=H\arabic*]
\item Reducing the processing fluency of a video with visual distortions makes it less credible.
\end{enumerate}

\hrulegray
\noindent
\textbf{Study~II} used a four-level design. Participants were randomly assigned to \textsc{reduced visual clarity}, \textsc{reduced speech clarity}, \textsc{audio--visual asynchrony}, or \textsc{baseline}. The study tested three hypotheses:
\begin{enumerate}[label=H\arabic*]
    \item Reducing the processing fluency of video with visual distortions makes it less credible (replicating Study~I).
    \item Reducing the processing fluency of the speech in a video with auditory distortions makes it less credible.
    \item Asynchrony between video and audio reduces the video's processing fluency and makes it less credible.
\end{enumerate}
\hrulegray
\noindent
\textbf{Study~III} used a four-level design. Participants were randomly assigned to \textsc{synthesized video}, \textsc{synthesized narration}, \textsc{synthesized video and narration}, or \textsc{baseline}. The study tested three hypotheses:
\begin{enumerate}[label=H\arabic*]
    \item A video with deepfake-generated visuals is processed less fluently than a recorded video and is less credible.
    \item A video with deepfake-generated narration is processed less fluently than one with recorded narration and is less credible.
    \item A video with deepfake-generated visuals and narration is processed less fluently than one with recorded narration and is less credible.
\end{enumerate}

\subsection{Preregistration}

All three studies were preregistered on the Open Science Framework along with their hypotheses, study plans, and statistical analysis plans (see Supplementary Materials).
All methods were carried out in accordance with the relevant guidelines and regulations for empirical research, and the experimental protocols were approved by the Research Ethics Committee at the IT University of Copenhagen.

\subsection{Participants}

All participants were US-based, English-speaking adults recruited via the online platform \textit{Prolific}\footnote{\url{https://prolific.com}}.

\subsubsection*{Study~I}

Three hundred and sixty US Americans were recruited in July 2023. After removal of $27$ participants not adhering to the preregistered quality criteria, data from $333$ participants were analyzed. The cohort comprised $174$ males, $147$ females, and $12$ participants who identified with other genders. Participants' mean age was $40.6$ years ($SD = 13.2$). Educational levels ranged from `High school or less' ($n=139$), `Bachelor' ($n=130$), `Master' ($n=56$), to `PhD' ($n=8$). The mean completion time was about four minutes ($M=$~\SI{234}{\second}$,\ SD = 114$).

\subsubsection*{Study~II}

In May 2023, we recruited US-based English-speaking participants via \textit{Prolific}. A pilot study with 16 participants indicated an effect size of Cohen's $f = 0.22$. We recruited $400$ participants (see~\nameref{S0_Appendix} for power analysis details). After exclusion of participants not meeting the preregistered quality criteria, data from $355$ participants were analyzed. The cohort comprised $187$ males, $156$ females, and $12$ participants who identified with other genders. Mean age was $43.0$ years ($SD = 12.9$). Educational levels ranged from `High school or less' ($n=150$), `Bachelor' ($n=154$), `Master' ($n=41$), to `PhD' ($n=10$). The mean completion time was slightly above 15 minutes ($M=$~\SI{980}{\second}$,\ SD = 365$).

\subsubsection*{Study~III}

In September 2023, we recruited $400$ US-based English-speaking participants via \textit{Prolific} (see~\nameref{S0_Appendix} for power analysis). After exclusion of participants not meeting the required quality criteria, data from 365 participants were analyzed. The cohort comprised 185 males, 170 females, and 10 participants who did not self-describe as either of those. Mean age was $44.4$ years ($SD = 16.5$). Educational levels ranged from `High school or less' ($n=129$), `Bachelor' ($n=177$), `Master' ($n=47$), to `PhD' ($n=12$). The mean completion time was about 17 minutes ($M=$~\SI{1005}{\second}$,\ SD = 363$).

\subsection{Dependent variables}

Across the three studies, we used self-report measures and multiple-choice knowledge tests; Study~I employed a subset of the measures used in Studies~II and~III (see details in~\autoref{tab:DependentMeasuresStudy1} and ~\autoref{tab:DependentMeasuresStudy2}).

\subsubsection*{Processing fluency}

Following \citet{graf_measuring_2018}, subjective processing fluency was measured with a slider, with internally stored responses between 0 (difficult) and 100 (easy). This measure was used in all three studies.

\subsubsection*{Message credibility}

Based on the scale by \citet{appelman_measuring_2016}, message credibility was computed as the mean of three 7-point Likert items \textit{accurate}, \textit{authentic}, and \textit{believable}, with response options ranging from 1 (``describes very poorly'') to 7 (``describes very well''). This scale was used in all three studies.

\subsubsection*{Digital alteration detection}

In all three studies, digital alteration detection was measured using the question ``Do you think the video has been digitally altered?'' with response options ``Yes'', ``No'', and ``I don't know''.

\subsubsection*{Learning}

In Study~II and Study~III, conceptual and factual learning were measured as the pre-to-post difference in a multiple choice questionnaire, following guidelines by \citet{petersen_pedagogical_2021}. Pre- and post-stimulus knowledge questionnaires were administered with mixed item order and embedded attention checks.

\subsubsection*{Source vividness, attentional focus, and liking}

In Study~II and Study~III, source vividness and attentional focus were measured on 7-point Likert scales anchored at ``strongly disagree'' and ``strongly agree''. Message credibility (as above) and liking were measured as 7-point Likert scales anchored between ``describes very poorly'' and ``describes very well''. For all multi-item scales, we report means.

\subsection{Stimuli}

\subsubsection*{Common stimulus video}

An educational video was produced to serve as a common stimulus source for all three studies. The video was \SI{447}{seconds} long and featured an actor presenting a script adapted from \citet{petersen_pedagogical_2021}. All video editing was performed in Adobe After Effects 23.2.1 (see \nameref{S1_Appendix} and \nameref{S2_Appendix} for details on video processing). The original script consisted of sections on general virology, measles, Zika, and COVID-19; we excluded the COVID-19 section to avoid interference from prevalent, non-scientific interpretations of the disease.

We chose the educational-video genre for four reasons. First, it allowed us to present complex, domain-specific information that is likely unfamiliar to most participants, thereby reducing the likelihood that credibility judgments could be resolved by simply validating content against prior knowledge. Second, the topic and presentation were designed to be ideologically neutral, limiting the influence of pre-existing attitudes or motivated reasoning on credibility assessments. Third, the format is familiar and ecologically valid for contemporary online platforms (e.g., YouTube, TikTok), where viewers routinely encounter informational videos. Finally, we used an unknown presenter against a neutral background to minimize extraneous cues such as source reputation, studio branding, or contextual set dressing that could independently affect credibility.

\subsubsection*{Study~I: Visual fluency manipulation}

\begin{figure}[h!]
\centering

\begin{subfigure}{0.3\textwidth}
  \includegraphics[width=\textwidth]{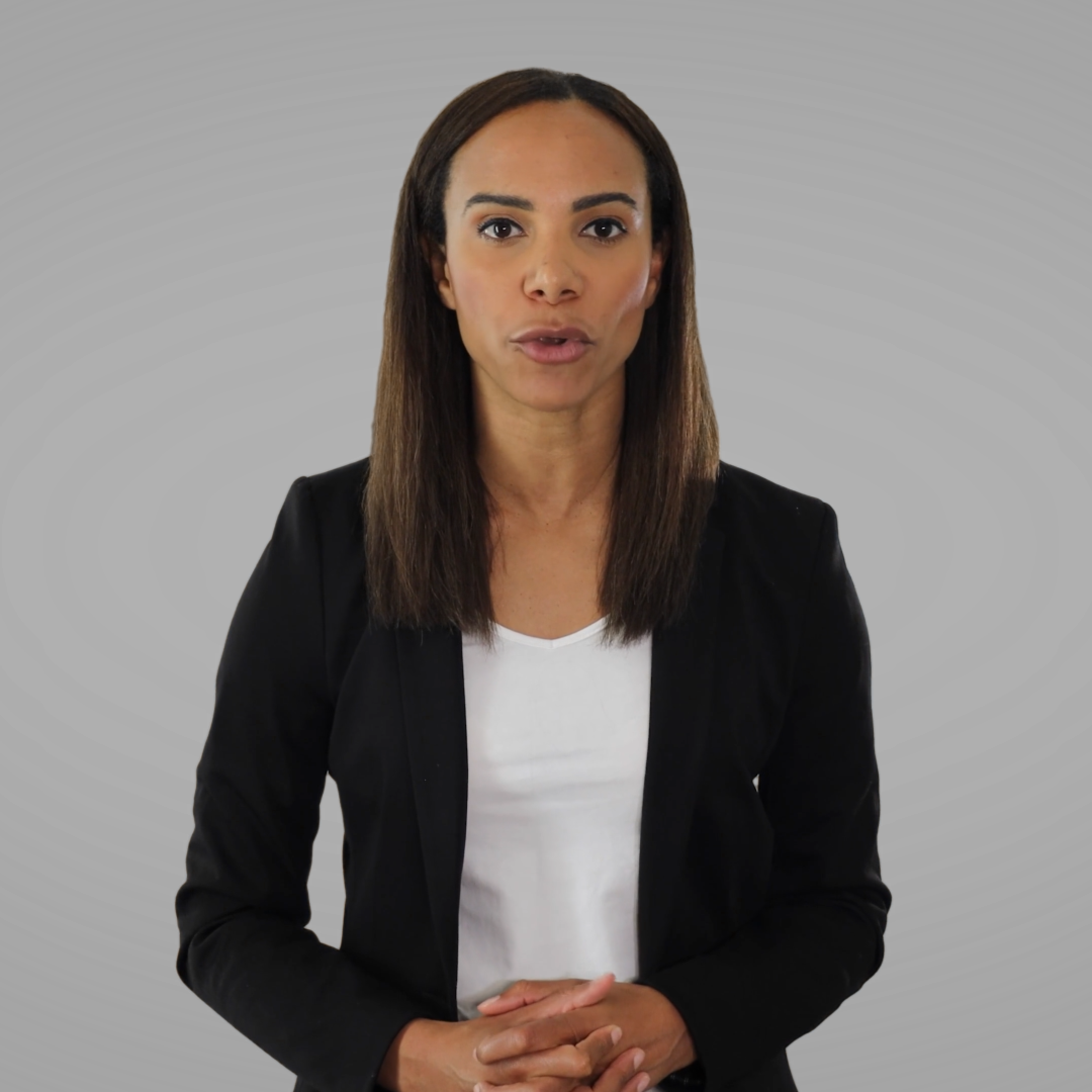}
  \caption{}
  \label{fig:Condition_A}
\end{subfigure}%
\hspace{1cm}%
\begin{subfigure}{0.3\textwidth}
  \includegraphics[width=\textwidth]{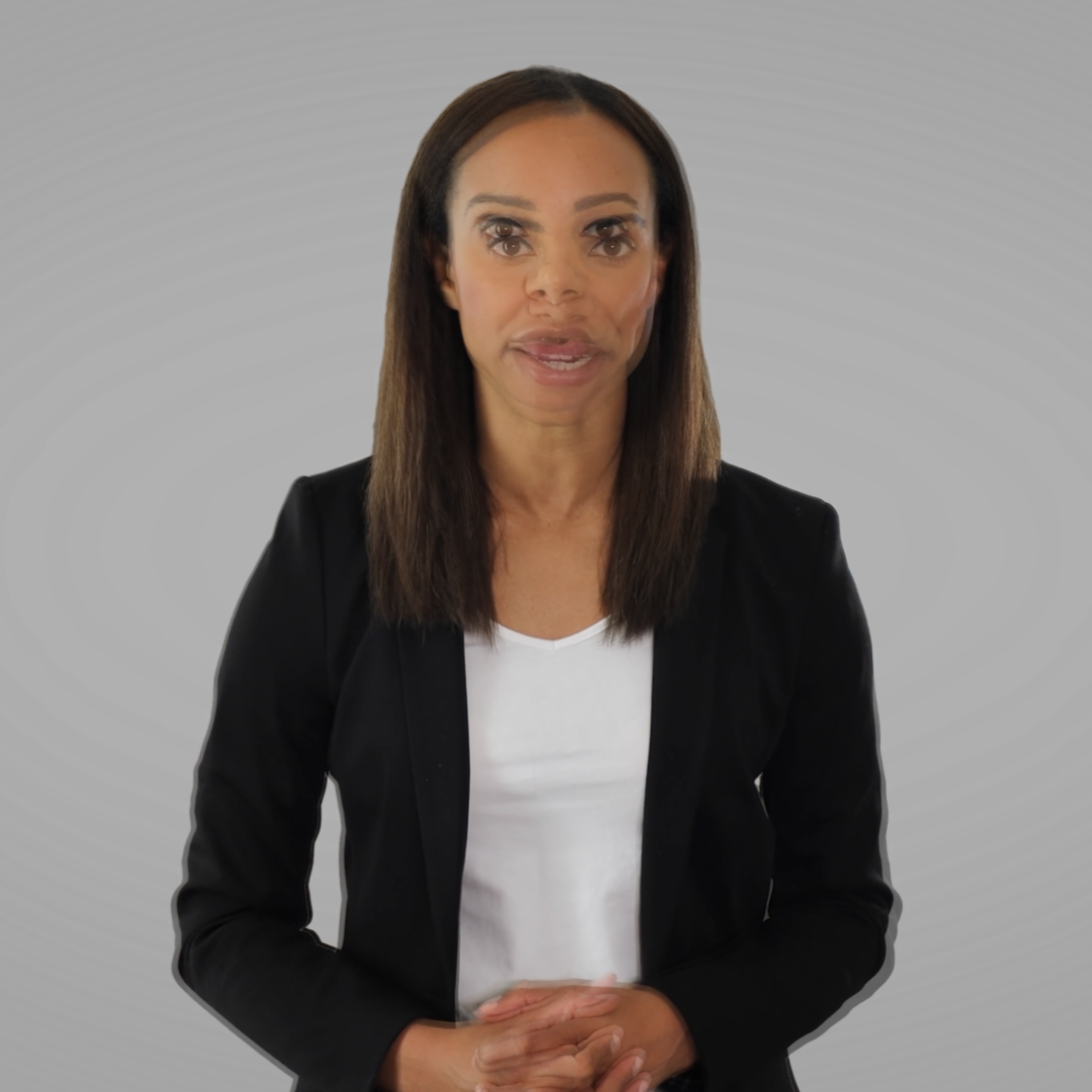}
  \caption{}
  \label{fig:Conditions_B}
\end{subfigure}
\caption{Stills of the two stimulus videos used in Study~I illustrating the differences between the experimental conditions: (a) a still from the unaltered, recorded video shown to participants in the \textsc{baseline} condition, and (b) a still from the distorted video shown to participants in the \textsc{reduced visual clarity} condition}. 
\Description{In two side-by-side images, the presenter of the stimulus video is shown in front of a gray, monochrome background. While the image (a) depicts her in focus, image (b) is a blend of two time-separated stills of the video and, thus, appears blurry where the presenter has moved between the two time-steps. The effect is particularly strong on facial features and hands.}
\label{fig:Conditions}
\end{figure}

In Study~I, we used the video's first section on general virology ($118$ seconds) as the source for the stimuli. In the \textsc{baseline} condition, the video was left unaltered. In the \textsc{reduced visual clarity} condition, the video featured a ghosting filter that superimposed a semitransparent duplicate of the original video over the original (see \autoref{fig:Conditions}). The speed of the superimposed video was manipulated to transition gradually between leading and lagging. These manipulations were chosen to distort the viewing experience without introducing elements that might be perceived as deceptive, thereby reducing the likelihood that credibility ratings were driven merely by suspicion.

\subsubsection*{Study~II: Visual, speech, and synchrony manipulations}

In Study~II, the full \SI{447}{seconds} video was used unchanged in the \textsc{baseline} condition and as the basis for all three treatment stimuli. As in Study~I, the \textsc{reduced visual clarity} condition featured a ghosting filter. Here, the ``ghost'' layer was only $30\%$ opaque, to achieve a less distorted viewing experience.

The \textsc{reduced speech clarity} condition applied a mild echo to the original narration. We selected this transformation to lower listening fluency while preserving all lexical content and timing, thereby minimizing the risk that credibility judgments were driven by suspicions of deliberate tampering rather than by audio quality per se (see \nameref{S2_Appendix} for details).
The \textsc{audio--visual asynchrony} stimulus featured a one-second delayed audio track that brought the speech and the presenter's lip movements out of alignment.

\subsubsection*{Study~III: Deepfake-based manipulations}

\begin{figure}[h!]
\centering

\begin{subfigure}{0.3\textwidth}
  \includegraphics[width=\textwidth]{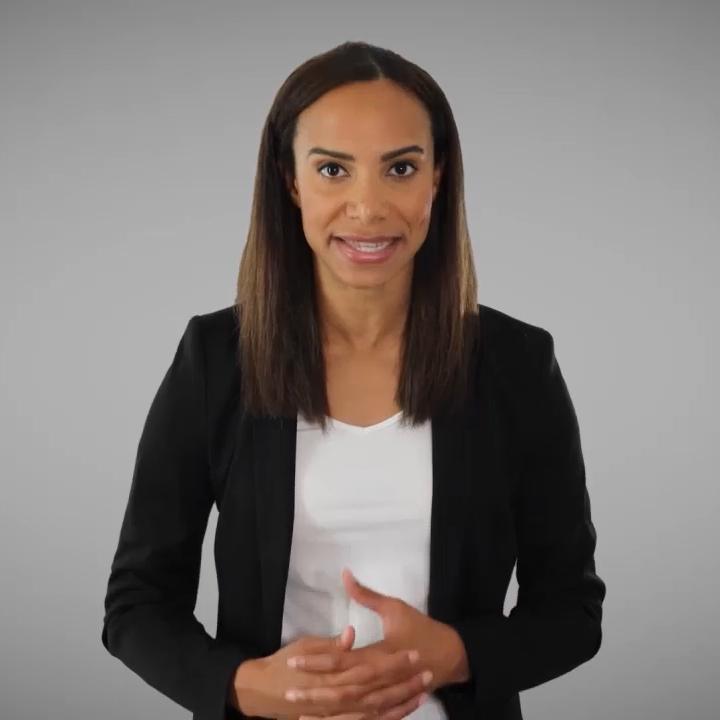}
  \caption{}
  \label{fig:StudyIII_Condition_A}
\end{subfigure}
\hspace{1cm}
\begin{subfigure}{0.3\textwidth}
  \includegraphics[width=\textwidth]{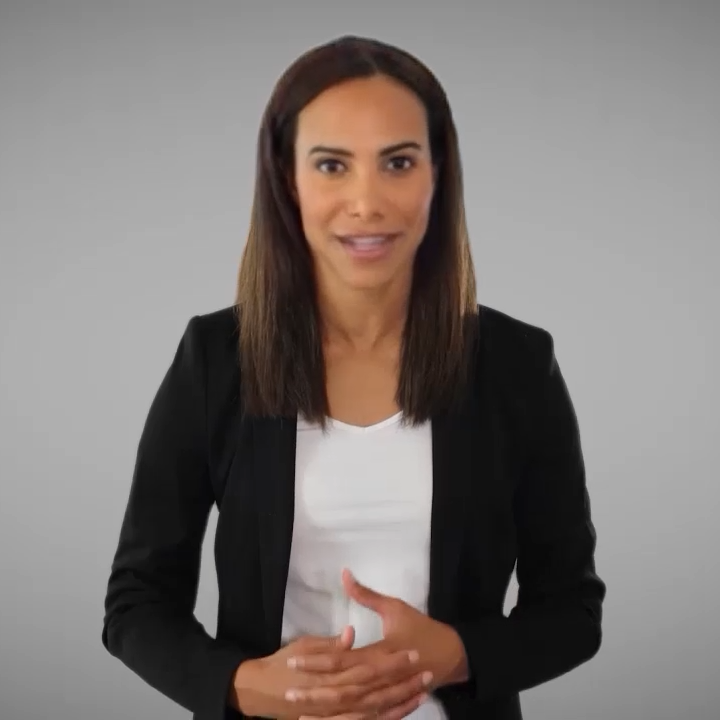}
  \caption{}
  \label{fig:StudyIII_Conditions_B}
\end{subfigure}
\caption{Stills showing the stimulus videos used in Study~III illustrating the visual differences between the experimental conditions: (a) a still from the unaltered, recorded video shown to participants in the conditions \textsc{baseline} and \textsc{synthesized narration}, and (b) a still from the deepfake video shown to participants in the \textsc{synthesized video} and \textsc{synthesized video and narration} conditions.}
\label{fig:StudyIII_Conditions}
\end{figure}

In Study~III, the same recorded educational video served as the \textsc{baseline} stimulus. Deepfake versions of the presenter's face and narration were created to produce the three experimental conditions (see \autoref{fig:StudyIII_Conditions}).

We used the Latent Image Animator with ``vox-pt'' weights by Wang and colleagues~\cite{wang_latent_2022} to synthesize a deepfake version of the video. This approach requires a $256\times256$ close-up video of a face and a $256\times256$ portrait picture to generate a $256\times256$ video of the portrait picture exhibiting the behavior captured in the video. The resulting animated portrait was then pasted into and aligned with the source video to obtain the deepfaked stimulus video.
We synthesized a version of the presenter's narration by fine-tuning a \texttt{So-Vits-Svc-fork} model\footnote{\url{https://github.com/voicepaw/so-vits-svc-fork}}. The model was trained on a \SI{14.25}{minutes} long recording of a diverse set of English sentences adapted from the Harvard sentences\footnote{\url{https://harvardsentences.com}}. Once trained and provided with the audio of the source video, the model synthesized a version of the original narration in the new voice.
The \textsc{synthesized video} stimulus was created by replacing the presenter's face in the source video with the deepfake animation while retaining the original audio. The \textsc{synthesized narration} stimulus replaced the source video's original narration track with the synthetic version. For the \textsc{synthesized video and narration} condition, we combined the deepfake stimulus video with the synthesized narration track.

\subsection{Procedure}

All studies were administered online using \textit{Gorilla}. In all three experiments, participants provided informed consent, confirmed audio readiness, reported demographics, and affirmed English proficiency. They then viewed their assigned stimulus video with playback controls disabled and completed attention checks. After the video, participants responded to the digital-alteration item and were debriefed.

\subsubsection*{Study~I}

Following video exposure, participants completed the processing-fluency and message-credibility measures (randomized order).

\subsubsection*{Studies II and III}

Studies~II and III followed the same core procedure but additionally included pre- and post-stimulus knowledge tests with mixed item order and attention checks. After viewing the video, participants completed processing-fluency and message-credibility measures as well as scales for source vividness, liking, and attentional focus (randomized order).

\section{Results}

Because most measures were non-normally distributed, we deviated from the preregistered plan in using conservative non-parametric tests (Kruskal–Wallis; Mann–Whitney $U$ for post hoc comparisons with Bonferroni correction where indicated). Effect sizes are reported as rank-biserial correlations. As preregistered for Study~III, Bayesian estimation complements the frequentist tests (see Appendix for full details).

\subsection{Study~I}

\subsubsection*{Message credibility}

Message credibility ratings (1--7) were higher in the \textsc{baseline} condition ($M = 6.2,\ SD = 0.8$) than in the \textsc{reduced visual clarity} condition ($M = 5.8,\ SD = 1.3$; see~\autoref{fig:ResultsStudy1-A}). The Mann--Whitney $U$ test indicated a significant difference between the conditions, $p < 0.0001$. The rank-biserial correlation was $0.50$.

\subsubsection*{Processing fluency}

\begin{figure*}[t]
\centering
\begin{subfigure}{.49\textwidth}
  \centering
  \includegraphics[width=\linewidth]{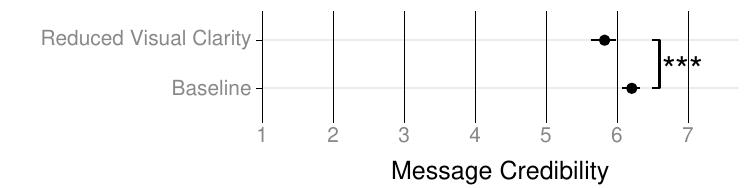}
  \caption{}
  \label{fig:ResultsStudy1-MessageCredibility}
\end{subfigure}
\begin{subfigure}{.49\textwidth}
  \centering
  \includegraphics[width=\linewidth]{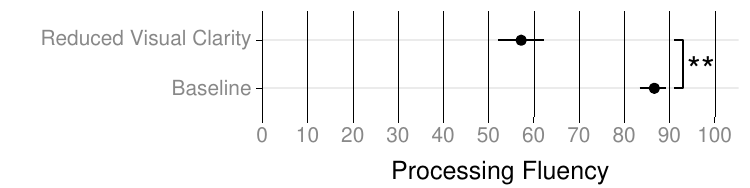}
  \caption{}
  \label{fig:ResultsStudy1-ProcessingFluency}
\end{subfigure}
\caption{Measurements of Message Credibility (a) and Processing Fluency (b) in Study~I divided by experimental condition. Dots represent mean values, horizontal lines indicate bootstrapped confidence intervals. Significant differences are indicated with brackets and asterisks: One asterisk (*) $p < 0.05$, two asterisks (**) $p < 0.01$, three asterisks (***) $p < 0.0001$.}
\Description{The figure presents two graphs comparing the effects of Reduced Visual Clarity and Baseline on (a) Message Credibility and (b) Processing Fluency, based on a study. In both graphs, dots represent the mean values of each condition, with horizontal lines showing bootstrapped confidence intervals.}
\label{fig:ResultsStudy1-A}
\end{figure*}

Processing fluency ratings averaged $86.7$ ($SD = 19.0$) in the \textsc{baseline} condition and $57.2$ ($SD = 34.2$) in the \textsc{reduced visual clarity} condition (see~\autoref{fig:ResultsStudy1-A}). The Mann--Whitney $U$ test indicated a significant difference between conditions, $p = 0.004$. The rank-biserial correlation was $0.18$.

\subsubsection*{Alteration detection}

\begin{figure}[h!]
  \begin{center}
    \includegraphics[width=0.75\textwidth]{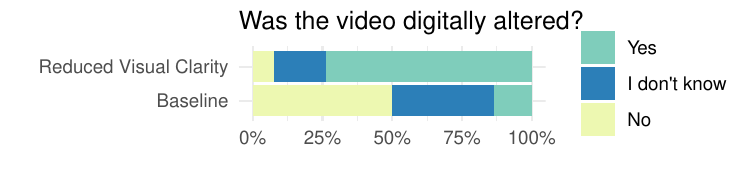}
  \end{center}
  \caption{Percentage of responses to the alteration detection question by treatment group in Study~I.}
  \Description{This figure presents a bar chart showing the percentage of responses to the question, ``Was the video digitally altered?'' for two groups: Reduced Visual Clarity and Baseline. The bars represent the percentage of participants who answered Yes (light gray), I don't know (medium gray), or No (dark gray). For the Reduced Visual Clarity group, the majority of participants answered ``I don't know'', followed by ``Yes'', with the smallest portion answering ``No''. For the Baseline group, most participants answered ``No'', with smaller portions answering ``I don't know'' and ``Yes''.}
  \label{fig:Study1_MD-Comparison}
\end{figure}

More participants detected a digital alteration in the \textsc{reduced visual clarity} condition ($n=126$) than in the \textsc{baseline} condition ($n=22$). More participants were uncertain about digital alterations in the \textsc{baseline} group ($n=59$) than in the \textsc{reduced visual clarity} group ($n=32$). Similarly, more participants in the \textsc{baseline} group did not detect an alteration ($n=81$) compared to the \textsc{reduced visual clarity} group ($n=13$; see~\autoref{fig:Study1_MD-Comparison}). Building on work by~\citet{altay2024people}, who found that the belief of text being AI-generated negatively impacting its perceived credibility, we expanded on our preregistered analysis and examined whether believing that the video was digitally altered was associated with lower credibility. A Spearman rank correlation indicated a moderate negative association between alteration detection and credibility ($r(331) = -.25, \, p < .001$), see \nameref{S4_Appendix} for details.

\subsubsection*{Exploratory analyses}

Exploratory correlations indicated no significant relationship between age and processing fluency ($r(331) = 0.06$, $p = 0.26$) but a small positive correlation between age and message credibility ($r(331) = 0.18$, $p = 0.0008$). Kruskal--Wallis tests did not show significant relationship between participant education and processing fluency ($p = 0.10$) or message credibility ($p = 0.76$).

\subsection{Study~II}

\subsubsection*{Message credibility}

A Kruskal--Wallis test demonstrated a significant difference in message credibility scores between conditions, $H(3) = 13.39$, $p = 0.004$. Subsequent pairwise Mann--Whitney $U$ tests with Bonferroni adjustment revealed significant differences between the \textsc{reduced speech clarity} condition ($M=6.22,\ SD = 1.05$) and the \textsc{reduced visual clarity} condition ($M = 5.87,\ SD = 1.01$); $p = 0.024$, rank-biserial correlation $= 0.24$, as well as between \textsc{reduced speech clarity} and \textsc{audio--visual asynchrony} ($M = 5.81,\ SD = 1.23$); $p = 0.019$, rank-biserial correlation $= 0.25$ (see~\autoref{fig:ResultsStudy2}).

\subsubsection*{Processing fluency}

A Kruskal--Wallis test demonstrated a significant overall difference in processing fluency scores between conditions, $H(3) = 9.36$, $p = 0.02$. However, the pairwise differences between conditions were not significant after Bonferroni correction (see~\autoref{fig:ResultsStudy2}).

\subsubsection*{Learning}

\begin{figure}[h!]
\begin{subfigure}{.49\textwidth}
  \centering
  \includegraphics[width=\linewidth]{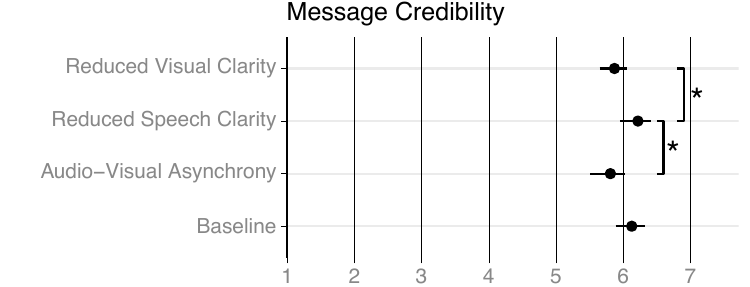}
  \caption{}
  \label{fig:ResultsStudy2-2}
\end{subfigure}
 \begin{subfigure}{.49\textwidth}
  \centering
  \includegraphics[width=\linewidth]{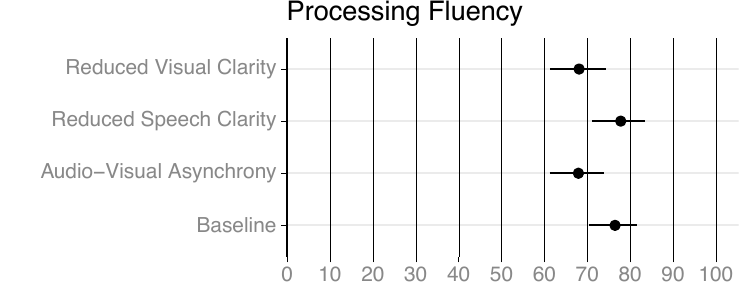}
  \caption{}
  \label{fig:ResultsStudy2-1}
\end{subfigure}
\newline
\begin{subfigure}{.49\textwidth}
  \centering
  \includegraphics[width=\linewidth]{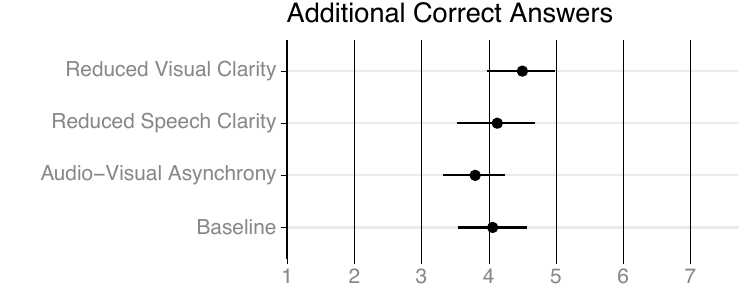}
  \caption{}
  \label{fig:ResultsStudy2-3}
\end{subfigure}
\begin{subfigure}{.49\textwidth}
  \centering
  \includegraphics[width=\linewidth]{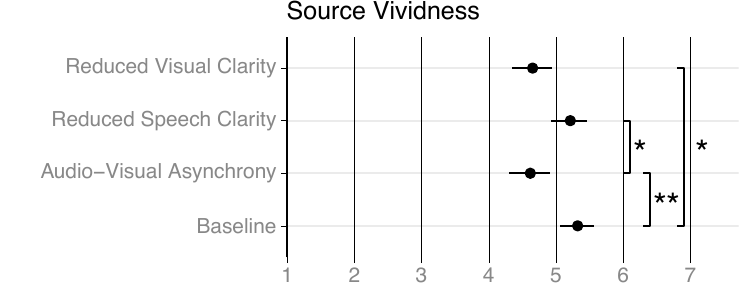}
  \caption{}
  \label{fig:ResultsStudy2-4}
\end{subfigure}
\caption{Measurements of Message Credibility (a), Processing Fluency (b), Factual and conceptual learning (c), and Source Vividness (d) in Study~II divided by experimental condition. Dots represent mean values, horizontal lines indicate bootstrapped confidence intervals. Significant differences according to Bonferroni-corrected tests are indicated with brackets and asterisks: One asterisk (*) $p < 0.05$, two asterisks (**) $p < 0.01$.}
\Description{The figure contains four graphs comparing the effects of four experimental conditions: ``Reduced Visual Clarity'', ``Reduced Speech Clarity'', ``Audio–Visual Asynchrony'', and ``Baseline''. The figures show (a) Message Credibility, (b) Processing Fluency, (c) Additional Correct Answers as measurement of Learning, and (d) Source Vividness. In graph (a), Message Credibility is highest in the Baseline condition, with lower scores in the other conditions. Significant differences are marked with asterisks. In graph (b), Processing Fluency values lie between 65 and 80 and show no significant pairwise differences after correction. In graph (c), Learning scores lie between 3.5 and 4.5, with no significant differences. In graph (d), Source Vividness values fall between 4.5 and 5.5, with significant differences between Baseline and some manipulated conditions.}
\label{fig:ResultsStudy2}
\end{figure}

Kruskal--Wallis tests did not reveal significant differences in factual learning ($H(3) = 5.65$, $p = 0.13$) or conceptual learning ($H(3) = 2.03$, $p = 0.57$) between conditions (see~\autoref{fig:ResultsStudy2}).

\subsubsection*{Source vividness}

A Kruskal--Wallis test indicated a statistically significant difference in source vividness scores across conditions, $H(3) = 17.12,\ p = 0.001$. Post hoc Mann--Whitney $U$ tests with Bonferroni adjustment showed significant differences between \textsc{baseline} ($M = 5.32,\ SD = 1.31$) and \textsc{reduced visual clarity} ($M = 4.65,\ SD = 1.44$); $p = 0.01$, rank-biserial correlation $= 0.26$, and between \textsc{baseline} and \textsc{audio--visual asynchrony} ($M = 4.62,\ SD = 1.45$); $p = 0.01$, rank-biserial correlation $= 0.28$. A significant difference was also observed between \textsc{audio--visual asynchrony} and \textsc{reduced speech clarity} ($M = 5.21,\ SD = 1.32$); $p = 0.03$, rank-biserial correlation $= 0.24$ (see~\autoref{fig:ResultsStudy2}).

\subsubsection*{Liking}

A Kruskal--Wallis test indicated a significant difference in liking across conditions, $H(3) = 10.22$, $p = 0.02$. Post hoc Mann--Whitney $U$ tests with Bonferroni adjustment revealed a significant difference between \textsc{baseline} ($M = 4.23,\ SD = 0.94$) and \textsc{audio--visual asynchrony} ($M = 3.83,\ SD = 0.89$); $p = 0.03$, rank-biserial correlation $= 0.24$.

\subsubsection*{Attentional focus}

A Kruskal--Wallis test did not uncover a significant difference in attentional focus across conditions, $H(3) = 6.37$, $p = 0.09$.

\subsubsection*{Alteration detection}

\begin{figure}[h!]
\begin{center}
     \includegraphics[width=0.75\textwidth]{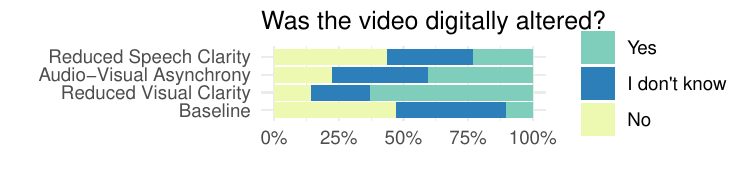}
  \end{center}
  \caption{Percentage of responses to the alteration detection question in Study~II by treatment group.}
  \Description{This figure displays a bar chart summarizing the responses to the question, ``Was the video digitally altered?'' across four treatment groups in Study~II: Reduced Speech Clarity, Audio–Visual Asynchrony, Reduced Visual Clarity, and Baseline. The bars represent the percentage of responses for each group, divided into three categories: Yes (light gray), I don't know (medium gray), No (dark gray).}
  \label{fig:Study2_MD-Comparison}
\end{figure}

All groups included participants who selected ``yes'' when asked whether the video was digitally altered. The share of ``yes'' responses was largest in the \textsc{reduced visual clarity} condition ($63\%$), followed by \textsc{audio--visual asynchrony} ($41\%$), \textsc{reduced speech clarity} ($23\%$), and \textsc{baseline} ($10\%$). The proportion answering ``I don't know'' was largest in the \textsc{baseline} condition ($43\%$), followed by \textsc{audio--visual asynchrony} ($37\%$), \textsc{reduced speech clarity} ($33\%$), and \textsc{reduced visual clarity} ($23\%$). This pattern and the lack of a substantial difference between baseline and manipulated conditions suggest generally high uncertainty regarding perceived video alteration that is most effectively resolved by overt visual alteration, but that persists despite the absence of any alteration (see~\autoref{fig:Study2_MD-Comparison}). Here too, a Spearman correlation indicated a negative association between alteration detection and credibility rating ($r(353) = -.27, \, p < .001$).

\subsection{Study~III}

\subsubsection*{Message credibility}

\begin{figure}[h!]
\begin{subfigure}{0.45\textwidth}
    \includegraphics[width=\linewidth]{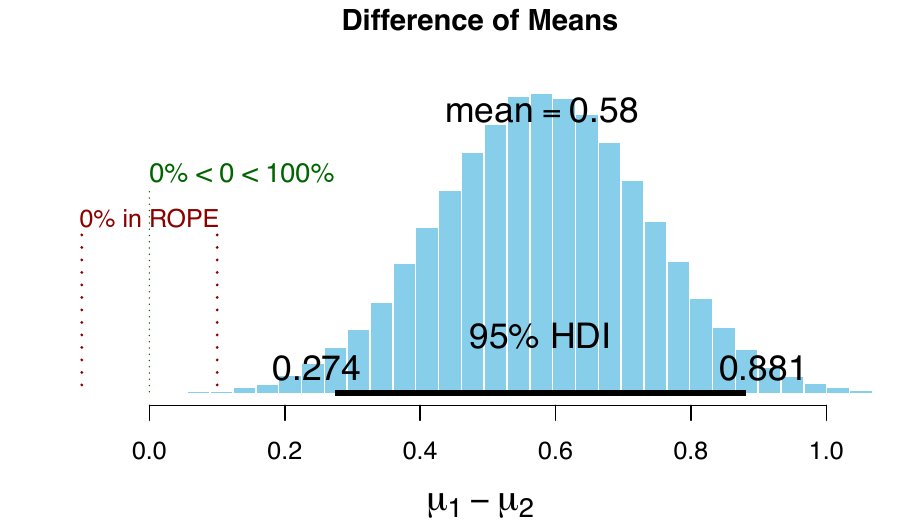}
    \caption{}
    \label{fig:BEST_MC_S3}
\end{subfigure}
\hfill
\begin{subfigure}{0.45\textwidth}
    \includegraphics[width=\linewidth]{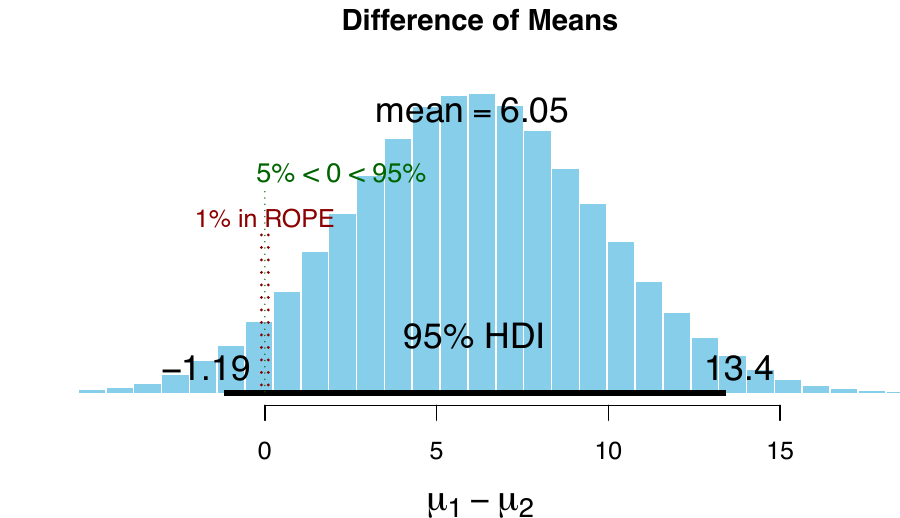}
    \caption{}
    \label{fig:BEST_PF_S3}
\end{subfigure}
\caption{Distribution of credible differences between the baseline video and the fully synthesized video. (a) shows the distribution of message credibility differences, (b) the distribution for processing fluency in Study~III. Bayesian estimation~\cite{kruschke_bayesian_2013} was used to calculate distributions of mean processing fluency and message credibility values for the \textsc{baseline} and \textsc{synthesized video and narration} conditions. The differences between the means are expressed in fractions of the pooled standard deviation (i.e., in units comparable to Cohen's $d$). The 95\% HDIs (High Density Intervals) indicate where the bulk of the credible differences fall. For processing fluency (b), the HDI is close to or includes 0, consistent with no meaningful difference between conditions. For message credibility (a), the HDI lies substantially above 0, and essentially all credible effect sizes exceed a small-effect threshold (Cohen's $d = 0.10$), indicating that the distance between the means is too large to accept $H_0$.}
\Description{This figure is comprised of two histograms depicting a roughly normal distribution each. Horizontal black bars marked as ``95\% HDI'' extend between the lower and upper tails of both histograms, with endpoints labelled with their corresponding values. The Message Credibility HDI extends from approximately 0.27 to 0.88, and the Processing Fluency HDI extends from approximately -1.19 to 13.4. Both subfigures feature an interval around the 0-point marked as ``ROPE'' (Region of Practical Equivalence). While the Message Credibility HDI does not overlap the ROPE, the Processing Fluency HDI extends over it.}
\label{fig:BESTStudy3}
\end{figure}

The manipulations significantly affected message credibility, $H(3) = 14.95$, $p = 0.002$. Post hoc Mann--Whitney $U$ tests with Bonferroni correction indicated a significant difference between \textsc{baseline} ($M=6.18,\ SD = 0.99$) and \textsc{synthesized video and narration} ($M=5.64,\ SD = 1.17$); $p = 0.02$, rank-biserial correlation $= 0.24$ (see~\autoref{fig:ResultsStudy3}). Bayesian estimation corroborated this effect: the distribution of credible differences in message credibility between \textsc{baseline} and \textsc{synthesized video and narration} was entirely above 0 and almost entirely outside a region of practical equivalence corresponding to Cohen's $d < 0.1$ (see~\autoref{fig:BESTStudy3}).

\subsubsection*{Processing fluency}

The Kruskal--Wallis test indicated no statistically significant difference in processing fluency scores across conditions, $H(3) = 2.35$, $p = 0.503$ (see~\autoref{fig:ResultsStudy3}). The Bayesian analysis yielded a 95\% highest density interval (HDI) for the difference between \textsc{baseline} and \textsc{synthesized video and narration} that included 0 and lay largely within the region of practical equivalence, supporting the conclusion that $H_0$ can be accepted for processing fluency (see~\autoref{fig:BESTStudy3}).

\subsubsection*{Learning}

\begin{figure}[h!]
\begin{subfigure}{.49\textwidth}
  \centering
  \includegraphics[width=\linewidth]{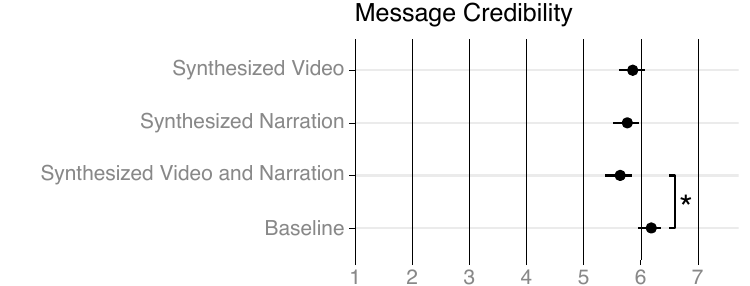}
  \caption{}
  \label{fig:MessageCredibility_S3}
\end{subfigure}
 \begin{subfigure}{.49\textwidth}
  \centering
  \includegraphics[width=\linewidth]{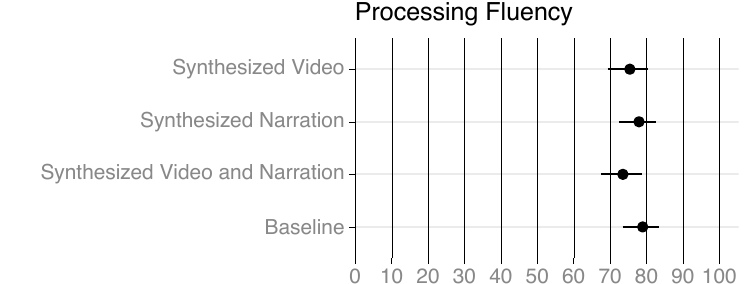}
  \caption{}
  \label{fig:ProcessingFluency_S3}
\end{subfigure}
\newline
\begin{subfigure}{.49\textwidth}
  \centering
  \includegraphics[width=\linewidth]{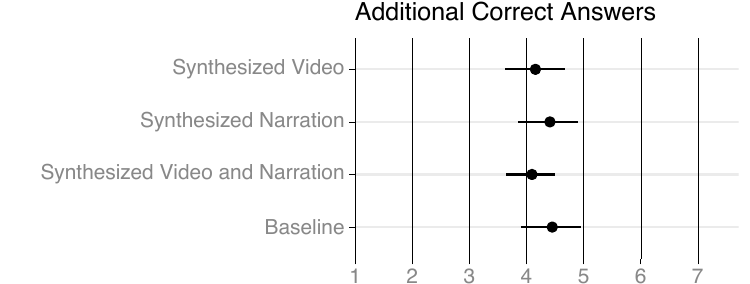}
  \caption{}
  \label{fig:Learning_S3}
\end{subfigure}
\begin{subfigure}{.49\textwidth}
  \centering
  \includegraphics[width=\linewidth]{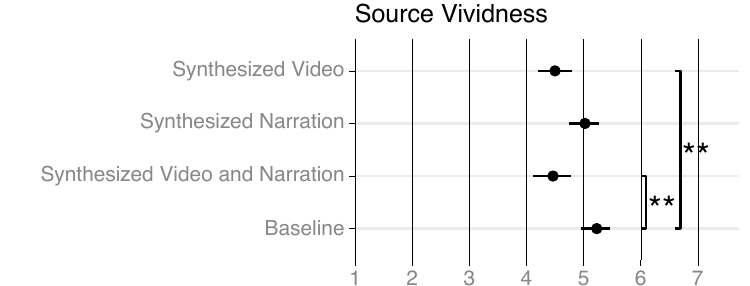}
  \caption{}
  \label{fig:SourceVividness_S3}
\end{subfigure}
\caption{Measurements of Message Credibility (a), Processing Fluency (b), Learning (factual and conceptual combined) (c), and Source Vividness (d) in Study~III divided by experimental condition. Dots represent mean values, horizontal lines indicate bootstrapped confidence intervals. Significant differences according to Bonferroni-corrected tests are indicated with brackets and asterisks: One asterisk (*): $p < 0.05$, two asterisks (**): $p < 0.01$.}
\Description{The figure contains four graphs comparing the effects of four experimental conditions: ``Synthesized Video'', ``Synthesized Narration'', ``Synthesized Video and Narration'', and ``Baseline''. The figure shows (a) Message Credibility, (b) Processing Fluency, (c) Additional Correct Answers as measurement of Learning, and (d) Source Vividness. In graph (a), Message Credibility is highest in the Baseline condition, with lower scores in the other conditions. The significant difference between Baseline and Synthesized Video and Narration is highlighted by an asterisk. In graph (b) the Processing Fluency values lie between 70 and 80 and show no significant differences. In graph (c), Learning scores lie between 4 and 5 on a scale of 1 to 7, with no significant differences. In graph (d), Source Vividness values fall between 4 and 5.5, with significant differences between Baseline and the conditions involving synthetic video.}
\label{fig:ResultsStudy3}
\end{figure}

Kruskal--Wallis tests indicated no significant differences in factual learning ($H(3) = 0.72$, $p = 0.87$) or conceptual learning ($H(3) = 2.55$, $p = 0.47$) between conditions (see~\autoref{fig:ResultsStudy3}). Bayesian estimation similarly indicated no practically meaningful differences in learning scores between the \textsc{baseline} and experimental conditions on either subscale.

\subsubsection*{Source vividness}

The manipulations significantly affected source vividness, $H(3) = 18.70$, $p = 0.0003$. Post hoc Mann--Whitney $U$ tests with Bonferroni correction showed significant differences between \textsc{baseline} ($M=5.23,\ SD = 1.29$) and \textsc{synthesized video and narration} ($p = 0.007$, rank-biserial correlation $= 0.28$), and between \textsc{baseline} and \textsc{synthesized video} ($p = 0.002$, rank-biserial correlation $= 0.31$; see~\autoref{fig:ResultsStudy3}).

\subsubsection*{Liking}

A Kruskal--Wallis test indicated no statistically significant differences in liking scores across conditions, $H(3) = 5.30$, $p = 0.15$.

\subsubsection*{Attentional focus}

A Kruskal--Wallis test indicated no statistically significant differences in attentional focus scores across conditions, $H(3) = 1.64$, $p = 0.65$.

\subsubsection*{Alteration detection}

\begin{figure}[h!]
\begin{center}
     \includegraphics[width=.75\textwidth]{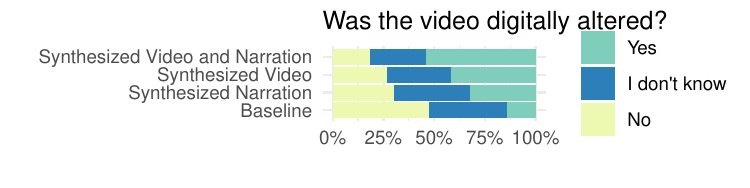}
\end{center}
  \caption{Percentage of responses to the alteration detection question in Study~III by treatment group.}
 \Description{This figure displays a bar chart summarizing the responses to the question, ``Was the video digitally altered?'' across four treatment groups in Study~III: Synthesized Video and Narration, Synthesized Video, Synthesized Narration, and Baseline. The bars represent the percentage of responses for each group, divided into three categories: Yes (light gray), I don't know (medium gray), No (dark gray).}
  \label{fig:Study3_MD-Comparison}
\end{figure}

All groups included participants who selected ``yes'' when asked whether the video was digitally altered. The share of ``yes'' responses was highest in the \textsc{synthesized video and narration} condition ($54\%$), followed by \textsc{synthesized video} ($42\%$), \textsc{synthesized narration} ($32\%$), and \textsc{baseline} ($14\%$). The proportion answering ``I don't know'' was similar in the \textsc{baseline} ($39\%$) and \textsc{synthesized narration} ($38\%$) conditions, and lower in \textsc{synthesized video} ($31\%$) and \textsc{synthesized video and narration} ($28\%$). The \textsc{synthesized video and narration} condition featured the fewest participants selecting ``no'' ($18\%$), and the \textsc{baseline} condition the highest ($47\%$). Thus, even in the absence of synthetic content, more than $10\%$ of participants reported detecting an alteration and roughly a third were uncertain. This uncertainty is lower in conditions with synthetic video, which also feature the highest rate of indicated detections. The effects of synthetic narration alone are comparatively limited (see~\autoref{fig:Study3_MD-Comparison}). Among the three studies the negative association between alteration detection and credibility rating was strongest here ($r(363) = -.40, \, p < .001$)

\section{Discussion}
Our study examined how visual and auditory manipulations affect perceived video credibility in the context of deepfakes. As summarized in \autoref{tab:study_synthesis}, the three studies form a cumulative progression from simple distortions to realistic deepfake artifacts. The results indicate that visual manipulations have a small but significant impact, though this effect cannot be explained by processing fluency. 

\subsubsection*{Video Credibility}
Our investigation into visual and auditory distortions and deepfake artifacts on video credibility across three studies reveals their significant influence on viewer perceptions. 

In Study~II, visual distortions and audio-video asynchrony reduced credibility compared to audio distortions alone but not the baseline, highlighting the role of sensory cues in credibility assessment. Study~III found that only the combination of deepfake visuals and narration significantly reduced credibility compared to the baseline, suggesting a small additive effect. Contrary to Newman and Schwarz~\cite{newman_good_2018}, we found no significant impact of audio manipulations on credibility.

\subsubsection*{Processing Fluency}
Our findings do not conclusively demonstrate the role of processing fluency in video perception. Study~I showed a correlation between processing fluency and credibility; this was not evident in Study~II and III. The data showed participants responded in a dichotomous way; participants positioned the indicator either towards 0 (left) or 100 (right) from its default position in the middle of the range.

\subsubsection*{Alteration Detection}
Groups of participants that indicated having and not having detected an alteration were present across all three studies, regardless of the presence of an manipulation. The data reveals a significant rate of inaccuracy in identifying distortions and deepfake videos, reflected in the high proportion of participants selecting ``I don't know.'' These findings are consistent with Vaccari and Chadwick's research, which links deepfakes to increased uncertainty~\cite{vaccari_deepfakes_2020}. Interestingly, the baseline condition in our studies highlights that the absence of manipulations does not reduce uncertainty. In Studies II and III, uncertainty was lowest in manipulation conditions and highest in baseline conditions without manipulations. Across studies, detecting (or believing to detect) an alteration was consistently associated with lower credibility. The negative correlations were small to moderate in Studies I and II and largest in Study~III, indicating that perceived alteration reduced credibility regardless of whether manipulations were actually present.

\subsubsection*{Additional Measures}
Our findings are in line with the association between source vividness and credibility as discussed by Lee and Shin in the context of deepfakes and misinformation~\cite{lee_something_2022}. We find, however, our source vividness measure to be more affected than our message credibility measure. This aligns with the understanding of video credibility assessment as complex and affected by multiple factors. 
While all participants learned from exposure to the stimulus, none of the employed manipulations affected their learning success. All learning-relevant information was mediated via the audio track of the videos. This suggests that the audio-based manipulations employed in Study~II and Study~III did not interfere with comprehensibility and that the measured reduction in source vividness and message credibility did not coincide with reduced apprehension. 

\begin{table}[h!]
  \centering
  \begin{tabularx}{\textwidth}{l Y Y Y}
    \toprule
    \textbf{Study} &
    \textbf{Manipulation} &
    \textbf{Key findings} &
    \textbf{Link to others} \\
    \midrule
    I &
    Visual distortion of a real recorded video. &
    Distortion sharply reduced processing fluency and message credibility,
    despite unchanged video content. &
    Establishes that \emph{non-probative} visual quality alone can lower
    credibility, motivating the broader manipulation space in II and III. \\
    \addlinespace
    II &
    Reduced visual clarity, reduced speech clarity,
    and audio--visual asynchrony. &
    Visual degradation and AV asynchrony lowered credibility and source vividness
    more than mild speech distortion; learning outcomes were unchanged. &
    Shows that visual and timing cues systematically affect credibility,
    and that these effects occur without harming learning; provides the bridge
    to ecologically realistic deepfake cues in III. \\
    \addlinespace
    III &
    Synthetic face and/or synthetic narration
    (deepfake video). &
    Fully synthetic face+voice reduced credibility and vividness most strongly.
    Visual deepfake artifacts drove larger credibility drops than synthetic
    narration; fluency and learning remained stable. &
    Demonstrates that the credibility penalties from I and II generalize to
    realistic deepfake artifacts, with strongest effects when both modalities
    are synthetic. Together, the studies trace a progression from abstract
    distortions to real-world deepfake cues. \\
    \bottomrule
  \end{tabularx}
  \caption{Summary of key findings across Studies~I--III on how visual distortions, audio distortions, and deepfake artifacts shape video credibility.}
  \label{tab:study_synthesis}
\end{table}

Beyond their substantive results, the studies also offer a methodological contribution for HCI. By experimentally decomposing synthetic-media manipulations into visual, auditory, and temporal components, and by separating deepfake-generated faces from deepfake-generated narration, we provide a reusable framework for probing how specific perceptual cues shape credibility judgments during interaction. This level of control is valuable for designing and evaluating transparency features, warning cues, provenance indicators, and other interface mechanisms that must target particular classes of artifacts rather than treat ``deepfakes'' as a single category. The open release of all stimuli, code, and preregistrations further supports iterative design and enables HCI researchers to build on well-specified, reproducible manipulations rather than ad-hoc media examples.

\subsection{Limitations}
This work has several limitations that qualify the scope of its claims. First, our samples were recruited via Prolific. Such panels tend to over represent relatively internet‐savvy participants. This may shift sensitivity to visual and audio irregularities compared to the general population. The direction of any bias is not obvious a priori: greater familiarity may heighten vigilance (increasing detection and decreasing credibility) or normalize minor artifacts (attenuating effects). Second, the studies were limited to U.S.-based, English-speaking participants. Cultural norms, media ecologies, and baseline trust in audiovisual media vary across societies; so may the salience of particular artifact types (e.g., eye contact norms, dubbing conventions). Our findings therefore warrant cross-cultural replications and extensions in other languages and cultures. Third, our manipulations necessarily cover only a small slice of the space of possible distortions and deepfake-generation pipelines. Different generation procedures likely produce different artifacts, and they evolve rapidly. We targeted representative classes of distortions, but did not and cannot comprehensively sample the design space. A further limitation is that each condition relied on a single video stimulus; this constrains generalizability and underscores the need for follow-up work using multiple stimuli.
Beyond these core constraints, ecological factors also limit generalizability. We intentionally used an unknown presenter and an educational talk-to-camera format to minimize confounds such as source reputation and political identity; effects could differ in news, entertainment, or highly partisan contexts where social cues dominate. Finally, our key metacognitive measure, the single-slider self-reports of processing fluency, may be coarse or susceptible to response heuristics, and alteration detection was measured with a single item. More sensitive instrumentation (e.g., response latencies, secondary-task load, eye movements, or psychophysiological indices) would help triangulate the underlying processes.

\subsection{Implications}
Our findings suggest, media literacy curricula should explicitly train attention to video-internal cues that our studies implicate as credibility-relevant: (i) local inconsistencies in shading/illumination and skin texture; (ii) facial geometry and boundary seams; and (iii) audio–visual timing mismatches and mouth–speech incongruence. Instruction should emphasize two guardrails: absence of visible artifacts is not evidence of authenticity, and artifact detection should be combined with source and context checks rather than used in isolation.

\section{Conclusion}
As deepfakes proliferate, it becomes more pressing to understand how humans process manipulated videos. Understanding how video production and synthesis relate to credibility evaluations, judgments of truth, and information recall can bolster us against the malicious spread of misinformation. The present paper takes a first stab at this and investigates how video aspects other than the language-based content affect their credibility. Through three studies, we show that distortions of the visual content of videos and deepfake artifacts can negatively affect video credibility. The reduced credibility effect can be observed across manipulations that may be interpreted as indicators of deceit, such as deepfake-synthesis suitable for impersonation, and distortions that are not associated with malicious intentions, like the reduced visual clarity manipulation seen in Study~I and Study~II. We did not find evidence to support the hypothesis that the relationship between a video's non-informational content and its credibility is based on processing fluency effects. More research of the cognitive processes involved in video perception and evaluation is needed, not least to assess the threat posed by deepfake videos. Our research marks a step forward in the nuanced understanding of video credibility in the age when online videos become an ever more central medium while the proliferation of deepfake technology undermines its veracity.

\section{Acknowledgments}
This research was supported by a Foundational Integrity Research award by Meta Research. In service of readability and conciseness, ChatGPT was utilized to copy-edit author generated content of this manuscript.

\bibliographystyle{unsrtnat}
\bibliography{references} 

\clearpage

\appendix

\renewcommand{\thetable}{S\arabic{table}}
\renewcommand{\thefigure}{S\arabic{figure}}

\setcounter{table}{0}
\setcounter{figure}{0}

\section{Appendix}

\subsection{Appendix 1}\label{S0_Appendix}

\paragraph{Power Analysis}
For Study~II and III, a power analysis conducted with \textit{G*Power} (ANOVA: fixed effects, omnibus, one-way; $f = 0.22$, $\alpha = 0.05$, power = $0.95$, number of groups = $4$) yielded a target sample size of $356$ participants. We recruited 400 participants to account for invalid responses.

\subsection{Appendix 2}\label{S1_Appendix}
\paragraph{The stimulus source.}
The actor was filmed in a resolution of $3840\times2160$ in front of a green-screen that was later replaced by a gray background. The resulting video was resized to $1280\times720$ pixels and compressed with variable bitrate (target = \SI{0.5}{\mega\bit\per\second}; max = \SI{5}{\mega\bit\per\second}). All editing of the stimulus source video has been carried out using the visual effects software Adobe After Effects 23.2.1.

\subsection{Appendix 3}\label{S2_Appendix}

\paragraph{Audio-Echo Effect.}
The echo effect was created with the built-in delay effect of the visual effects software Adobe After Effects 23.2.1 (delay time: \SI{0.5}{\second}, delay amount: 25\%, feedback: 25\%, dry out: 25\%, wet out: 50\%).

\subsection{Appendix 4}
\begin{table}[h!]
\small
\def\arraystretch{1.1}
\begin{tabularx}{\textwidth}{p{2.5cm} p{4cm} p{4cm} p{3.25cm}}
\toprule
    \textbf{Dependent Variable} & \textbf{Instrument} & \textbf{Type} & \textbf{Prompt} \\
\midrule
    Subjective Processing Fluency & Single item scale adapted from \citet{graf_measuring_2018} & Slider anchored with ``difficult'' and ``easy'' & ``The process of watching this video was...'' \\ 
\midrule
    \multirow[t]{3}{4cm}{Message Credibility} & 
    \multirow[t]{3}{4cm}{Message credibility scale adapted from Appelman \& Sundar~\cite{appelman_measuring_2016}} &
    \multirow[t]{3}{4cm}{7-point Likert-type items anchored with ``describes very poorly'' and ``describes very well''} & 
    How well does the word `accurate' describe the video? \\
\cmidrule{4-4} 
     &  &  & How well does the word `authentic' describe the video? \\ 
\cmidrule{4-4} 
     &  &  & How well does the word `believable' describe the video? \\ 
\midrule
    Digital Manipulation Detection & Multiple choice item adapted from \citet{nightingale_can_2017} & Multiple choice question with the answer options: ``Yes'', ``I don't know'', and ``No'' & ``Do you think the video has been digitally altered?'' \\
\bottomrule
\end{tabularx}%
\caption{Dependent measures used in Study~I.}
\label{tab:DependentMeasuresStudy1}
\Description{The table shows the measures used in study 1, divided by the variable, who first introduced it, the type of measure, and how we asked participants about it.}
\end{table}

\begin{table}[h!]
\small
\begin{tabularx}{\textwidth}{p{4cm} p{5cm} X}
\toprule
    \textbf{Dependent Variable} & \textbf{Instrument} & \textbf{Type} \\
\midrule
    Conceptual Learning & Conceptual learning scale adapted from \citet{petersen_pedagogical_2021} & Eight multiple choice questions, with four answer options each, applied before and after stimulus presentation \\
\midrule
    Factual Learning & Factual learning scale adapted from \citet{petersen_pedagogical_2021} & Eight multiple choice questions, with four answer options each, applied before and after stimulus presentation \\ 
\midrule
    Subjective Processing Fluency & Single item scale adapted from \citet{graf_measuring_2018} & Slider anchored with ``difficult'' and ``easy'' \\
\midrule
    Message Credibility & Message credibility scale adapted from \citet{appelman_measuring_2016} & Three 7-point Likert-type items anchored with ``describes very poorly'' and  ``describes very well'' \\
\midrule
    Source Vividness & Source Vividness scale adapted from \citet{lee_something_2022} & Four 7-point Likert-type items anchored with ``strongly disagree'' and ``strongly agree'' \\
\midrule
    Liking & Online news liking scale adapted from \citet{sundar_exploring_1999} & Five 7-point Likert-type items anchored with ``describes very poorly'' and ``describes very well'' \\
\midrule
    Attentional Focus & Attentional Focus scale adapted from \citet{busselle_measuring_2009} & Three 7-point Likert-type items anchored with ``strongly disagree'' and ``strongly agree'' \\
\midrule
    Digital Manipulation Detection & Multiple choice item adapted from \citet{nightingale_can_2017} & Multiple choice question with the answer options: ``Yes'', ``I don't know'', and ``No'' \\
\bottomrule
\end{tabularx}%
\caption{Dependent measures employed in Study~II and Study~III.}
\label{tab:DependentMeasuresStudy2}
\Description{The table shows the measures used in study 2, divided by the variable, who first introduced it, and the type of measure.}
\end{table}

\clearpage

\subsection{Appendix 5: Frequentist analysis}
\begin{table}[h]
\centering
    \begin{tabular}{lcc}
    \hline
    Test               & p-value & Cohen's d \\ \hline
    Processing Fluency & < 0.0001 & 1.06      \\ \hline
\end{tabular}
\caption{Study~I: Summary of T-test Results for Processing Fluency}
\label{S1_Table}
\end{table}


\begin{table}[h]
\centering
\begin{tabular}{lcc}
\hline
Test               & p-value & Cohen's d \\ \hline
Processing Fluency & = 0.0006 & 0.38      \\ \hline
\end{tabular}
\caption{Study~I: Summary of T-test Results for Message Credibility}
\label{S2_Table}
\end{table}


\begin{table}[h]
\centering
\begin{tabular}{lccccc}
\hline
Source & Df & Sum Sq & Mean Sq & F value & Pr(>F) \\ \hline
Condition & 3 & 7458 & 2486.2 & 2.844 & 0.0377* \\
Residuals & 351 & 306806 & 874.1 & & \\ \hline
\end{tabular}
\caption{Study~II: ANOVA Results for Processing Fluency Across Conditions}
\label{S3_Table}
\end{table}


\begin{table}[h]
\centering
\begin{tabular}{lc}
\hline
Comparison & P-value \\ \hline
Reduced Visual Clarity vs. Baseline & 0.35 \\
Audi-Visual Incoherence vs. Baseline & 0.34 \\
Reduced Auditive Clarity vs. Baseline & 1.00 \\
Reduced Auditive Clarity vs. Reduced Visual Clarity & 0.17 \\
Audi-Visual Incoherence vs. Reduced Visual Clarity & 1.00 \\
Audi-Visual Incoherence vs. Reduced Auditive Clarity & 0.17 \\ \hline
\end{tabular}
\caption{Study~II: Pairwise Comparisons for Processing Fluency Across Conditions. P-values adjusted using Bonferroni method}
\label{S4_Table}
\end{table}


\begin{table}[h]
\centering
\begin{tabular}{lccccc}
\hline
Source & Df & Sum Sq & Mean Sq & F value & Pr(>F) \\ \hline
Condition & 3 & 10.3 & 3.450 & 2.954 & 0.0326* \\
Residuals & 351 & 410.0 & 1.168 & & \\ \hline
\end{tabular}
\caption{Study~II: ANOVA Results for Message Credibility Across Conditions}
\label{S5_Table}
\end{table}


\begin{table}[h]
\centering
\begin{tabular}{lc}
\hline
Comparison & P-value \\ \hline
Reduced Visual Clarity vs. Baseline & 0.677 \\
Audi-Visual Incoherence vs. Baseline & 0.313 \\
Reduced Auditive Clarity vs. Baseline & 1.000 \\
Reduced Auditive Clarity vs. Reduced Visual Clarity & 0.190 \\
Audi-Visual Incoherence vs. Reduced Visual Clarity & 1.000 \\
Audi-Visual Incoherence vs. Reduced Auditive Clarity & 0.075 \\ \hline
\end{tabular}
\caption{Study~II: Pairwise Comparisons for Message Credibility Across Conditions (P-values adjusted using Bonferroni method)}
\label{S6_Table}
\end{table}


\begin{table}[h]
\centering
\begin{tabular}{lccccc}
\hline
Source & Df & Sum Sq & Mean Sq & F value & Pr(>F) \\ \hline
Condition & 3 & 21 & 6.992 & 1.694 & 0.168 \\
Residuals & 351 & 1448 & 4.126 & & \\ \hline
\end{tabular}
\caption{Study~II: ANOVA Results for Factual Learning Across Conditions}
\label{S7_Table}
\end{table}


\begin{table}[h]
\centering
\begin{tabular}{lccccc}
\hline
Source & Df & Sum Sq & Mean Sq & F value & Pr(>F) \\ \hline
Condition & 3 & 4.6 & 1.55 & 0.787 & 0.502 \\
Residuals & 351 & 691.3 & 1.97 & & \\ \hline
\end{tabular}
\caption{Study~II: ANOVA Results for Conceptual Learning Across Conditions}
\label{S8_Table}
\end{table}


\begin{table}[h]
\centering
\begin{tabular}{lccccc}
\hline
Source & Df & Sum Sq & Mean Sq & F value & Pr(>F) \\ \hline
Condition & 3 & 36.0 & 11.998 & 6.478 & 0.000281*** \\
Residuals & 351 & 650.1 & 1.852 & & \\ \hline
\end{tabular}
\caption{Study~II: ANOVA Results for Condition on Source Vividness}
\label{S9_Table}
\end{table}


\begin{table}[h]
\centering
\begin{tabular}{lc}
\hline
Comparison & P-value \\ \hline
Reduced Visual Clarity vs. Baseline & 0.0066 \\
Audi-Visual Incoherence vs. Baseline & 0.0040 \\
Reduced Auditive Clarity vs. Baseline & 1.0000 \\
Reduced Auditive Clarity vs. Reduced Visual Clarity & 0.0372 \\
Audi-Visual Incoherence vs. Reduced Visual Clarity & 1.0000 \\
Audi-Visual Incoherence vs. Reduced Auditive Clarity & 0.0239 \\ \hline
\end{tabular}
\caption{Study~II: Pairwise Comparisons for Source Vividness Across Conditions (P-values adjusted using Bonferroni method)}
\label{S10_Table}
\end{table}


\begin{table}[h]
\centering
\begin{tabular}{lccccc}
\hline
Source & Df & Sum Sq & Mean Sq & F value & Pr(>F) \\ \hline
Condition & 3 & 10.03 & 3.344 & 3.761 & 0.0111* \\
Residuals & 351 & 312.09 & 0.889 & & \\ \hline
\end{tabular}
\caption{Study~II: ANOVA Results for Condition on Liking}
\label{S11_Table}
\end{table}


\begin{table}[h]
\centering
\begin{tabular}{lc}
\hline
Comparison & P-value \\ \hline
Reduced Visual Clarity vs. Baseline & 0.151 \\
Audi-Visual Incoherence vs. Baseline & 0.031 \\
Reduced Auditive Clarity vs. Baseline & 1.000 \\
Reduced Auditive Clarity vs. Reduced Visual Clarity & 0.414 \\
Audi-Visual Incoherence vs. Reduced Visual Clarity & 1.000 \\
Audi-Visual Incoherence vs. Reduced Auditive Clarity & 0.102 \\ \hline
\end{tabular}
\caption{Study~II: Pairwise Comparisons for Liking Across Conditions (P-values adjusted using Bonferroni method)}
\label{S12_Table}
\end{table}


\begin{table}[h]
\centering
\begin{tabular}{lccccc}
\hline
Source & Df & Sum Sq & Mean Sq & F value & Pr(>F) \\ \hline
Condition & 3 & 8.9 & 2.976 & 1.046 & 0.372 \\
Residuals & 351 & 998.3 & 2.844 & & \\ \hline
\end{tabular}
\caption{Study~II: ANOVA Results for Condition on Attentional Focus}
\label{S13_Table}
\end{table}


\begin{table}[h]
\centering
\begin{tabular}{lccccc}
\hline
Source & Df & Sum Sq & Mean Sq & F value & Pr(>F) \\ \hline
Condition & 3 & 1679 & 559.7 & 0.823 & 0.482 \\
Residuals & 361 & 245374 & 679.7 & & \\ \hline
\end{tabular}
\caption{Study~III: ANOVA Results for Processing Fluency Across Conditions}
\label{S14_Table}
\end{table}


\begin{table}[h]
\centering
\begin{tabular}{lccccc}
\hline
Source & Df & Sum Sq & Mean Sq & F value & Pr(>F) \\ \hline
Condition & 3 & 15.1 & 5.040 & 4.229 & 0.00589** \\
Residuals & 361 & 430.2 & 1.192 & & \\ \hline
\end{tabular}
\caption{Study~III: ANOVA Results for Message Credibility Across Conditions}
\label{S15_Table}
\end{table}


\begin{table}[h]
\centering
\begin{tabular}{lc}
\hline
Comparison & P-value \\ \hline
RecVid-SynAud vs. RecVid-RecAud & 0.0582 \\
SynVid-RecAud vs. RecVid-RecAud & 0.2727 \\
SynVid-SynAud vs. RecVid-RecAud & 0.0044 \\
SynVid-RecAud vs. RecVid-SynAud & 1.0000 \\
SynVid-SynAud vs. RecVid-SynAud & 1.0000 \\
SynVid-SynAud vs. SynVid-RecAud & 1.0000 \\ \hline
\end{tabular}
\caption{Study~III: Pairwise Comparisons for Message Credibility Across Conditions (P-values adjusted using Bonferroni method)}

\label{S16_Table}
\end{table}


\begin{table}[h]
\centering
\begin{tabular}{lccccc}
\hline
Source & Df & Sum Sq & Mean Sq & F value & Pr(>F) \\ \hline
Condition & 3 & 40.3 & 13.424 & 6.739 & 0.000196*** \\
Residuals & 361 & 719.1 & 1.992 & & \\ \hline
\end{tabular}
\caption{Study~III: ANOVA Results for Source Vividness Across Conditions}
\label{S17_Table}
\end{table}


\begin{table}[h]
\centering
\begin{tabular}{lc}
\hline
Comparison & P-value \\ \hline
RecVid-SynAud vs. RecVid-RecAud & 1.0000 \\
SynVid-RecAud vs. RecVid-RecAud & 0.0032 \\
SynVid-SynAud vs. RecVid-RecAud & 0.0015 \\
SynVid-RecAud vs. RecVid-SynAud & 0.0797 \\
SynVid-SynAud vs. RecVid-SynAud & 0.0458 \\
SynVid-SynAud vs. SynVid-RecAud & 1.0000 \\ \hline
\end{tabular}
\caption{Study~III: Pairwise Comparisons for Source Vividness Across Conditions (P-values adjusted using Bonferroni method)}

\label{S18_Table}
\end{table}


\begin{table}[h]
\centering
\begin{tabular}{lccccc}
\hline
Source & Df & Sum Sq & Mean Sq & F value & Pr(>F) \\ \hline
Condition & 3 & 5.45 & 1.8161 & 2.183 & 0.0896 \\
Residuals & 361 & 300.26 & 0.8318 & & \\ \hline
\end{tabular}
\caption{Study~III: ANOVA Results for Liking Across Conditions}
\label{S19_Table}
\end{table}


\begin{table}[h]
\centering
\begin{tabular}{lccccc}
\hline
Source & Df & Sum Sq & Mean Sq & F value & Pr(>F) \\ \hline
Condition & 3 & 2.5 & 0.8494 & 0.314 & 0.815 \\
Residuals & 361 & 977.3 & 2.7072 & & \\ \hline
\end{tabular}
\caption{Study~III: ANOVA Results for Attentional Focus Across Conditions}
\label{S20_Table}
\end{table}
\clearpage

\subsection{Appendix 6: Bayesian analysis}

\begin{figure}[h]
\centering
\includegraphics[width=0.5\textwidth]{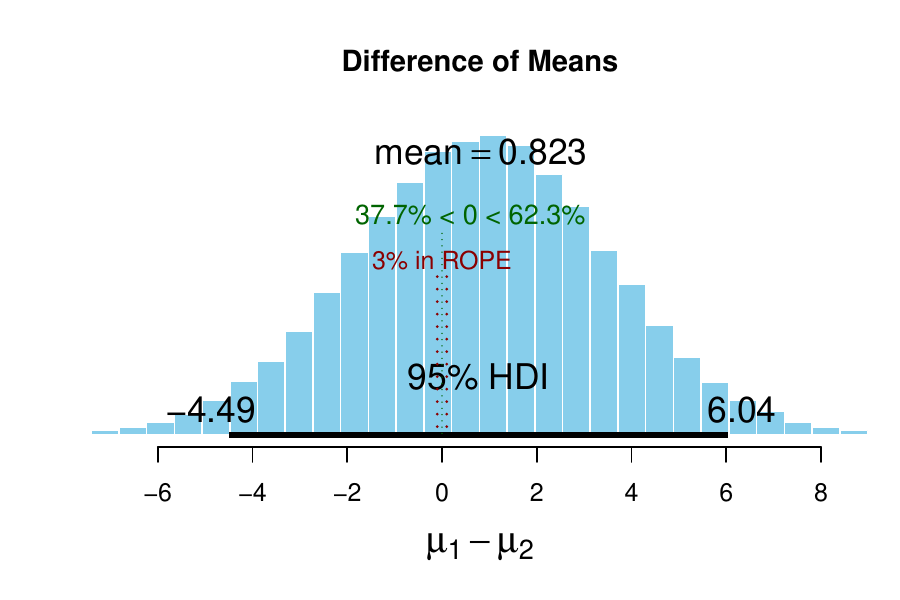}
\caption{Distribution of credible differences between the mean processing fluency of the baseline video and the synthesized narration video.}
\label{S1_Fig}
\Description{This figure shows a histogram showing the distribution of credible differences in processing fluency between baseline and synthesized narration.}
\end{figure}


\begin{figure}[h]
\centering
\includegraphics[width=0.5\textwidth]{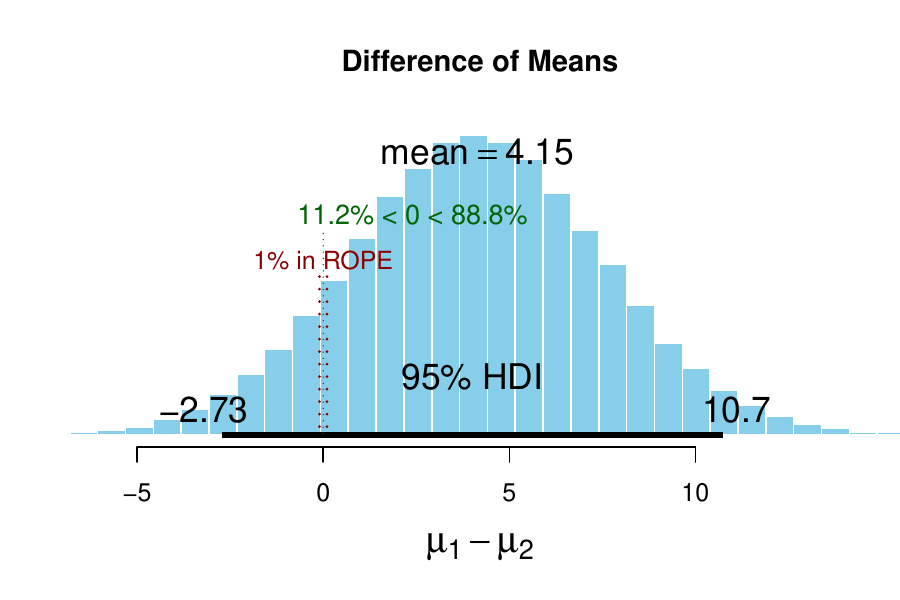}
\caption{Distribution of credible differences between the mean processing fluency of the baseline video and the synthesized video.}
\label{S2_Fig}
\Description{This figure shows a histogram showing the distribution of credible differences in processing fluency between baseline and synthesized video.}
\end{figure}


\begin{figure}[h]
\centering
\includegraphics[width=0.5\textwidth]{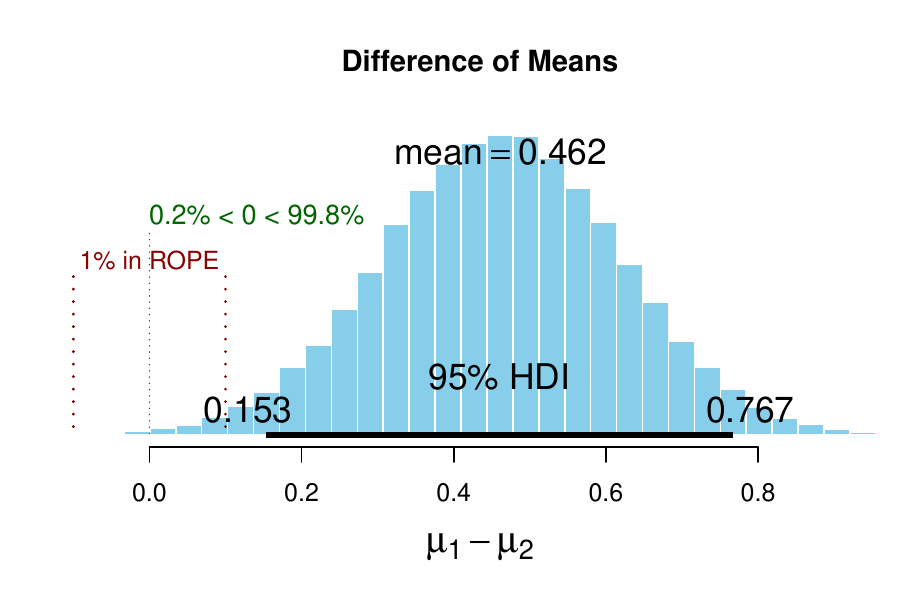}
\caption{Distribution of credible differences between the mean message credibility of the baseline video and the synthesized narration video.}
\Description{This figure shows a histogram showing the distribution of credible differences in message credibility between baseline and synthesized narration video.}
\label{S3_Fig}
\end{figure}


\begin{figure}[h]
\centering
\includegraphics[width=0.5\textwidth]{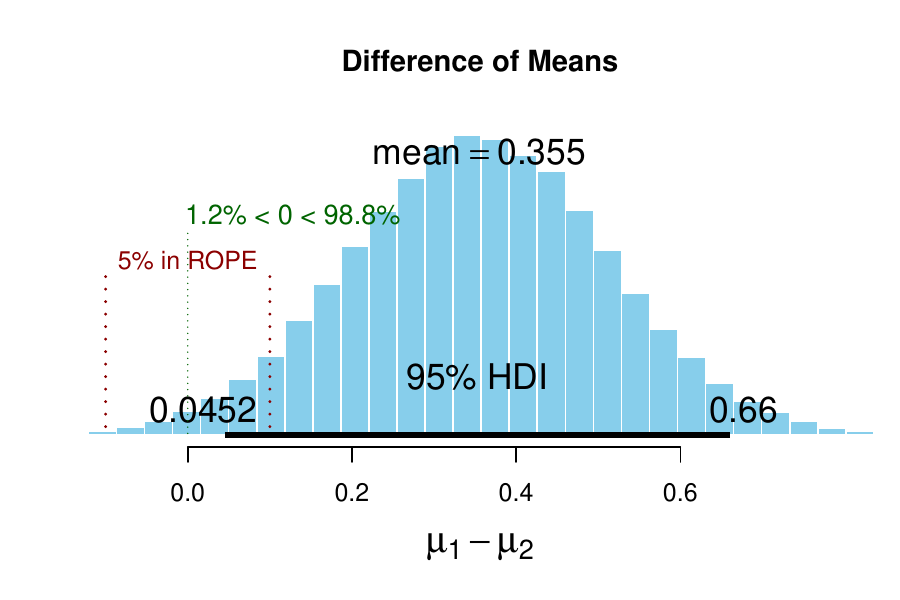}
\caption{Distribution of credible differences between the mean message credibility of the baseline video and the synthesized video.}
\Description{This figure shows a histogram showing the distribution of credible differences in message credibility between baseline and synthesized video.}
\label{S4_Fig}
\end{figure}


\begin{figure}[h]
\centering
\includegraphics[width=0.5\textwidth]{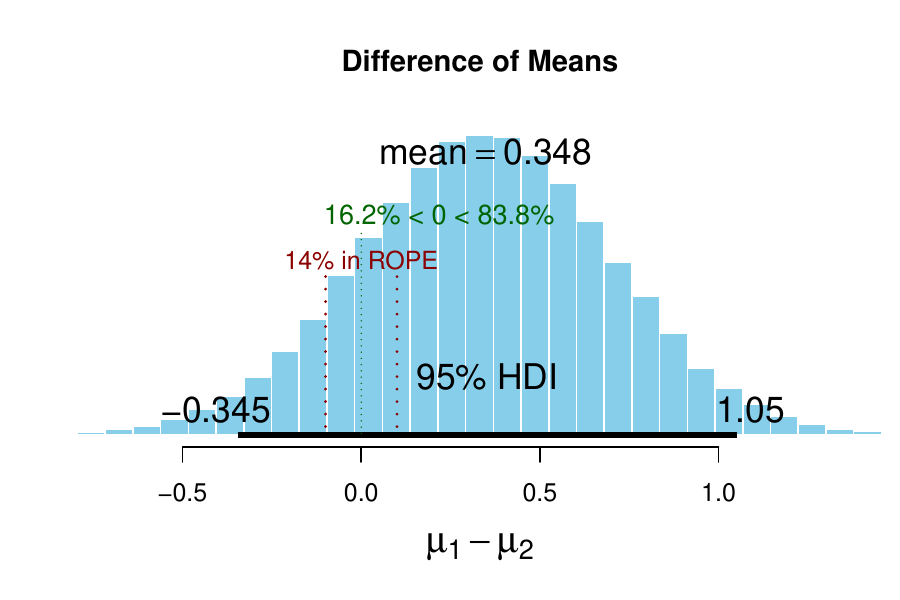}
\caption{Distribution of credible differences between the mean correct answers (Learning) of the baseline video and synthesized video and narration condition.}
\Description{This figure shows a histogram showing the distribution of credible differences in learning between baseline and synthesized video and narration condition.}
\label{S5_Fig}
\end{figure}


\begin{figure}[h]
\centering
\includegraphics[width=0.5\textwidth]{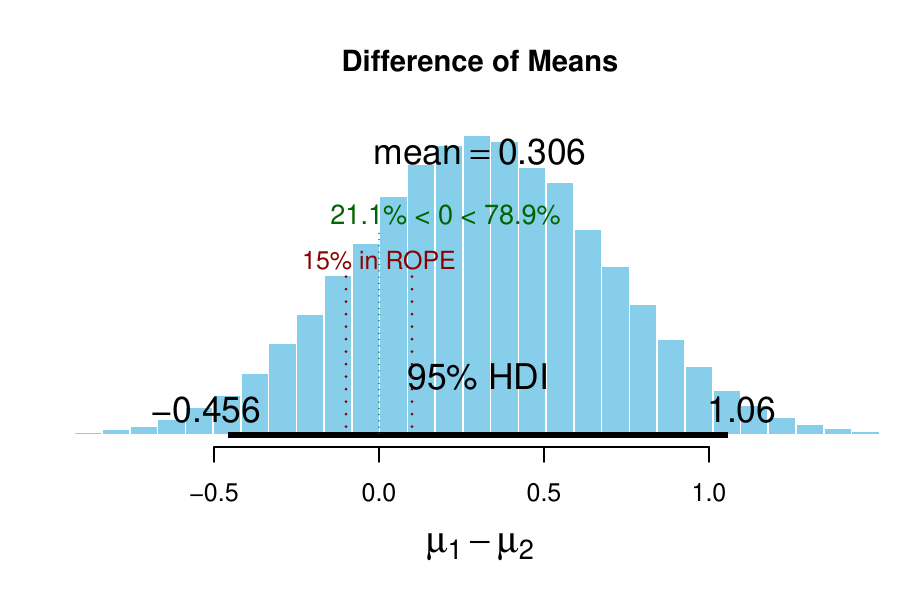}
\caption{Distribution of credible differences between the mean correct answers (Learning) of the baseline video and synthesized video condition.}
\Description{This figure shows a histogram showing the distribution of credible differences in learning between baseline and synthesized video condition.}
\label{S6_Fig}
\end{figure}


\begin{figure}[h]
\centering
\includegraphics[width=0.5\textwidth]{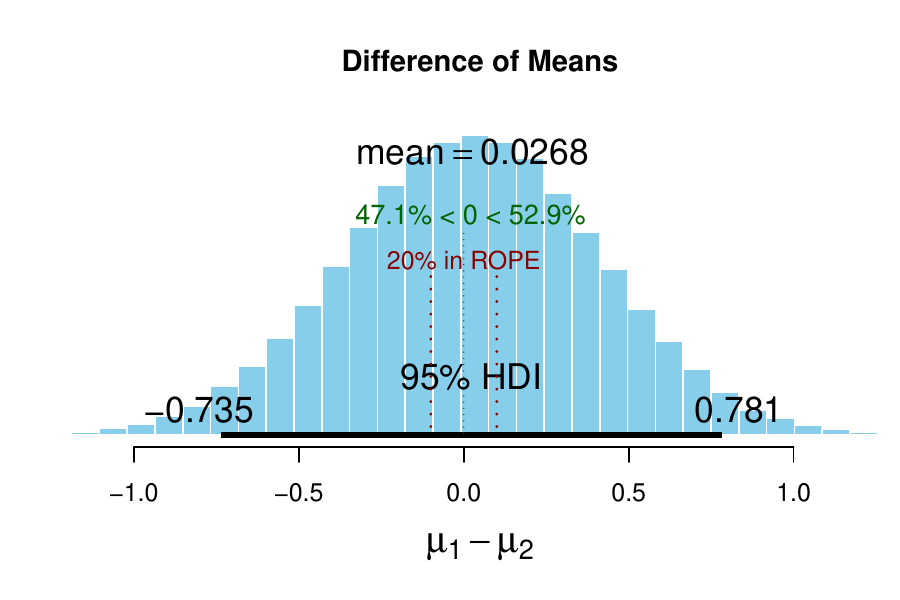}
\caption{Distribution of credible differences between the mean correct answers (Learning) of the baseline video and synthesized narration condition.}
\Description{This figure shows a histogram showing the distribution of credible differences in learning between baseline and synthesized narration condition.}
\label{S7_Fig}
\end{figure}


\begin{figure}[h]
\centering
\includegraphics[width=0.5\textwidth]{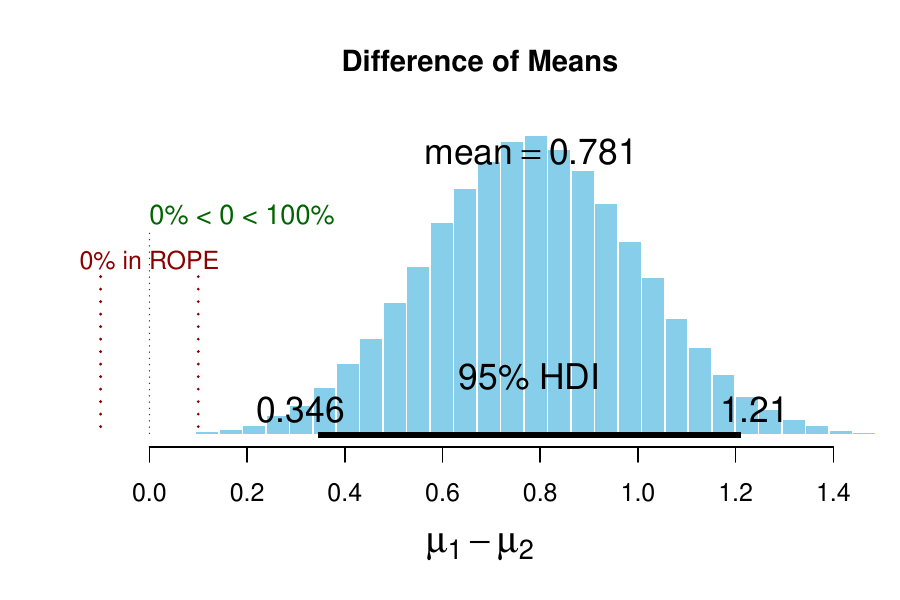}
\caption{Distribution of credible differences between the mean source vividness of the baseline video and synthesized video and narration condition.}
\Description{This figure shows a histogram showing the distribution of credible differences in vividness between baseline and synthesized video and narration condition.}
\label{S8_Fig}
\end{figure}


\begin{figure}[h]
\centering
\includegraphics[width=0.5\textwidth]{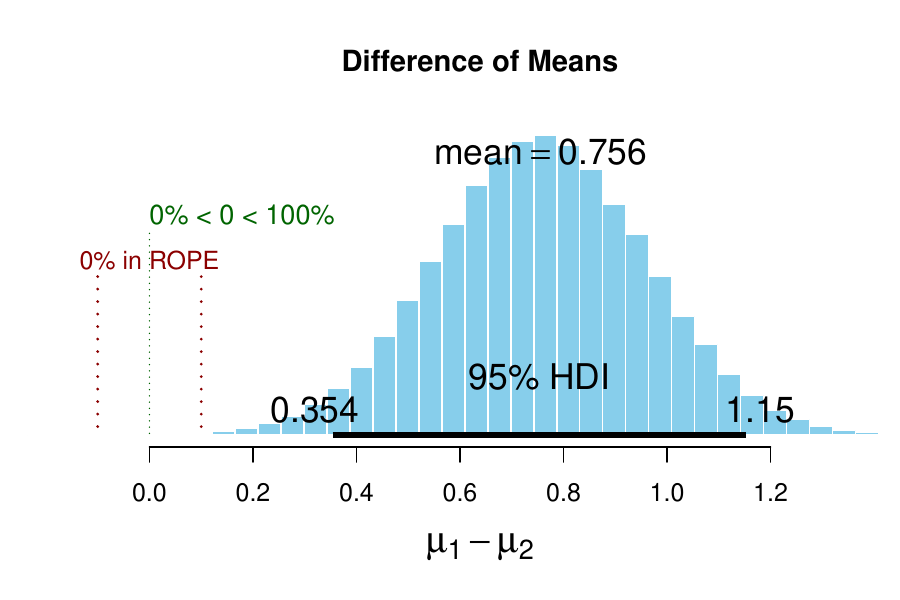}
\caption{Distribution of credible differences between the mean source vividness of the baseline video and synthesized video condition.}
\Description{This figure shows a histogram showing the distribution of credible differences in vividness between baseline and synthesized video condition.}
\label{S9_Fig}
\end{figure}


\begin{figure}[h]
\centering
\includegraphics[width=0.5\textwidth]{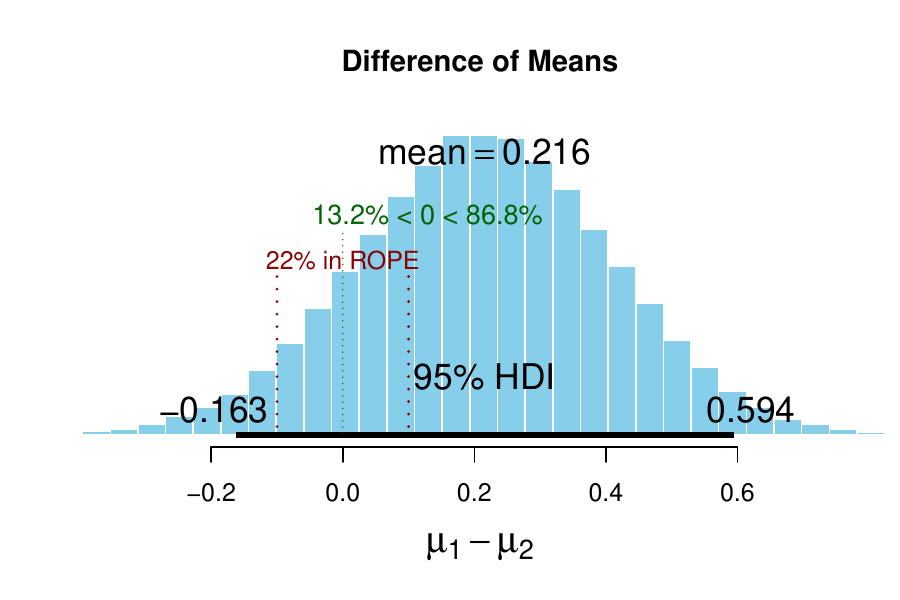}
\caption{Distribution of credible differences between the mean source vividness of the baseline video and synthesized narration condition.}
\Description{This figure shows a histogram showing the distribution of credible differences in vividness between baseline and synthesized narration condition.}
\label{S10_Fig}
\end{figure}


\begin{figure}[h]
\centering
\includegraphics[width=0.5\textwidth]{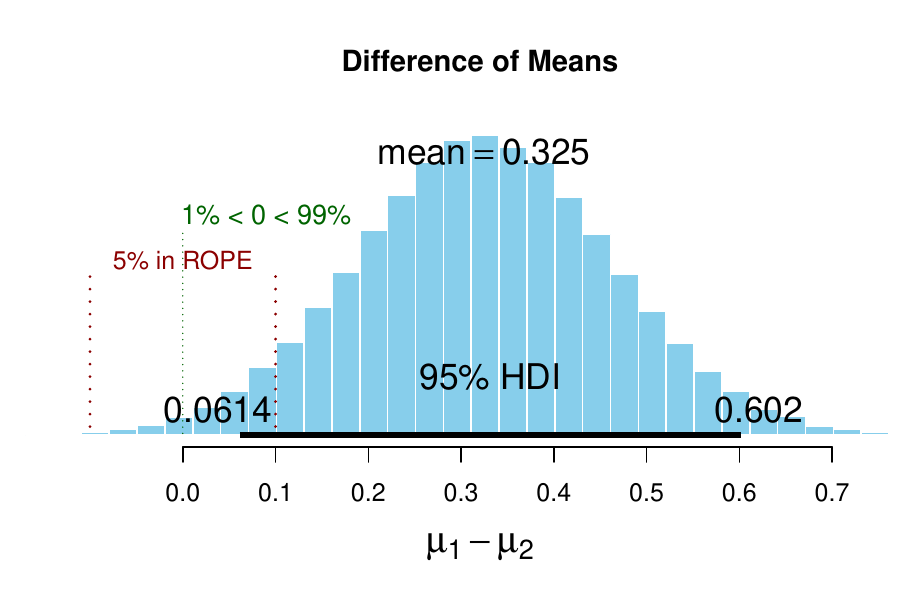}
\caption{Distribution of credible differences between the mean liking measurement of the baseline video and synthesized video and narration condition.}
\Description{This figure shows a histogram showing the distribution of credible differences in liking between baseline and synthesized video and narration condition.}
\label{S11_Fig}
\end{figure}


\begin{figure}[h]
\centering
\includegraphics[width=0.5\textwidth]{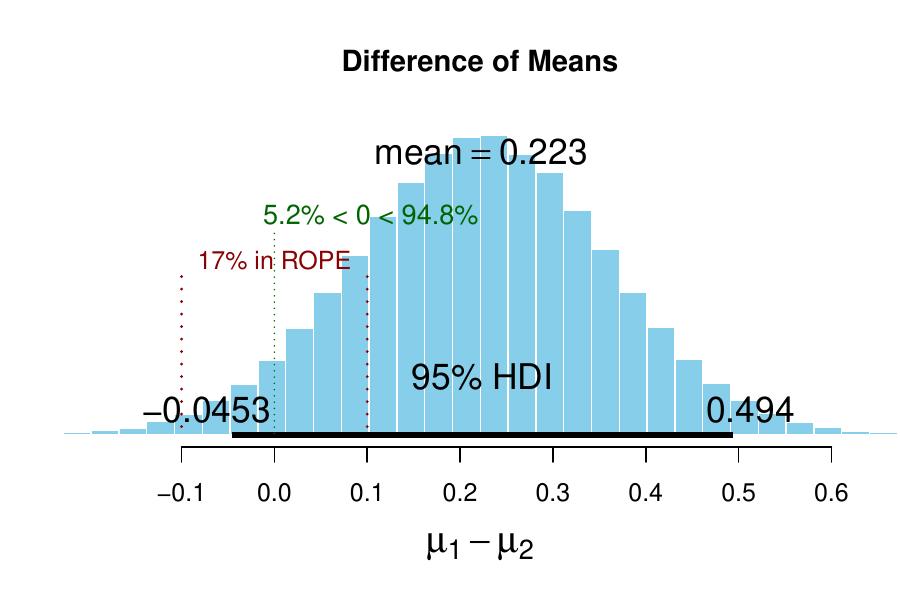}
\caption{Distribution of credible differences between the mean liking measurement of the baseline video and synthesized video condition.}
\Description{This figure shows a histogram showing the distribution of credible differences in liking between baseline and synthesized video condition.}
\label{S12_Fig}
\end{figure}


\begin{figure}[h]
\centering
\includegraphics[width=0.5\textwidth]{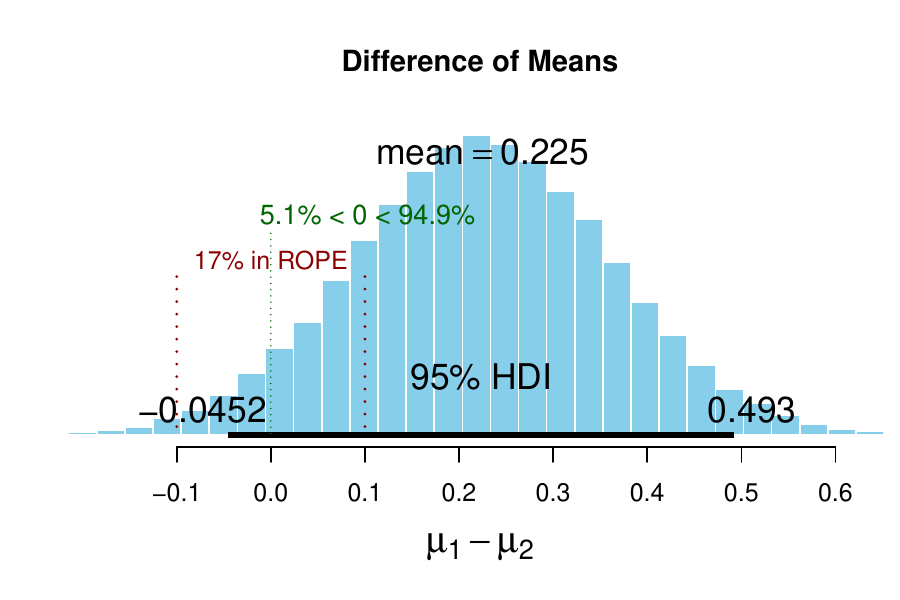}
\caption{Distribution of credible differences between the mean liking measurement of the baseline video and synthesized narration condition.}
\Description{This figure shows a histogram showing the distribution of credible differences in liking between baseline and synthesized narration condition.}
\label{S13_Fig}
\end{figure}


\begin{figure}[h]
\centering
\includegraphics[width=0.5\textwidth]{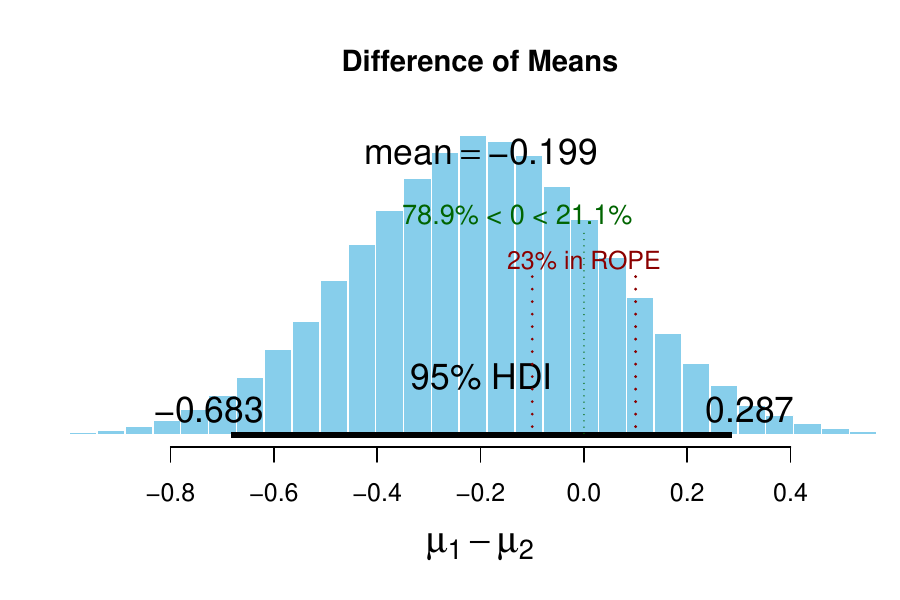}
\caption{Distribution of credible differences between the mean attentional focus of the baseline video and synthesized video and narration condition.}
\Description{This figure shows a histogram showing the distribution of credible differences in attention between baseline and synthesized video and narration condition.}
\label{S14_Fig}
\end{figure}


\begin{figure}[h]
\centering
\includegraphics[width=0.5\textwidth]{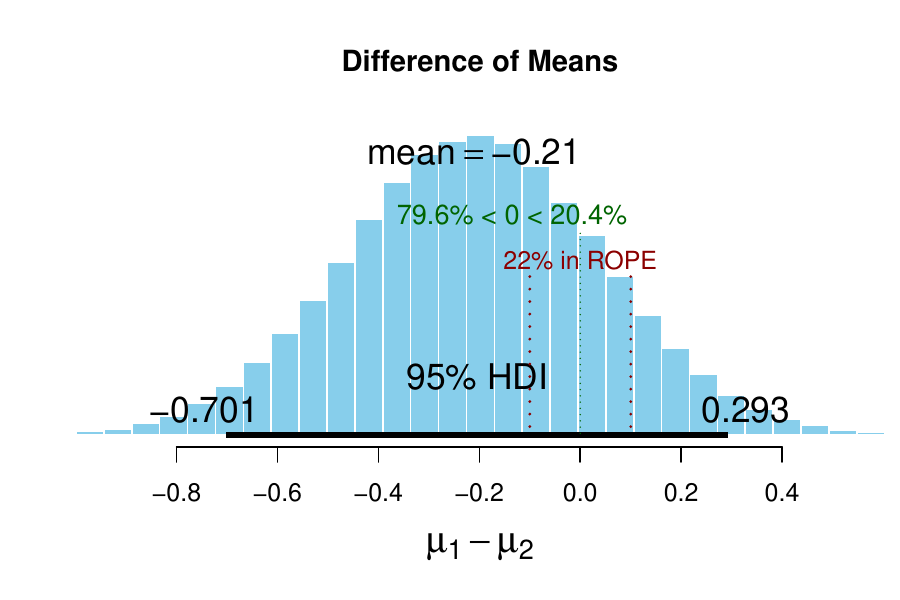}
\caption{Distribution of credible differences between the mean attentional focus of the baseline video and synthesized video condition.}
\Description{This figure shows a histogram showing the distribution of credible differences in attention between baseline and synthesized video condition.}
\label{S15_Fig}
\end{figure}


\begin{figure}[h]
\centering
\includegraphics[width=0.5\textwidth]{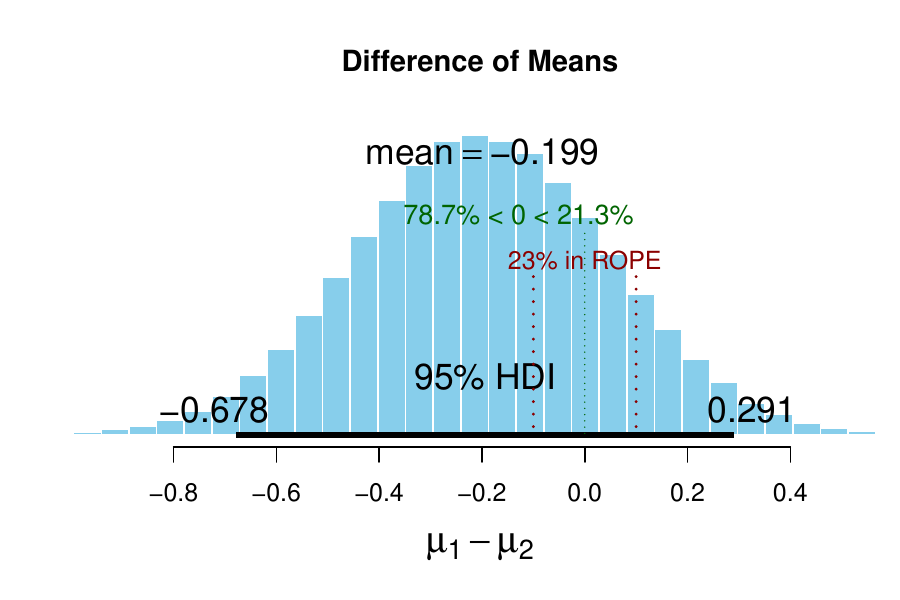}
\caption{Distribution of credible differences between the mean attentional focus of the baseline video and synthesized narration condition.}
\Description{This figure shows a histogram showing the distribution of credible differences in attention between baseline and synthesized narration condition.}
\label{S16_Fig}
\end{figure}

\clearpage
\subsection{Appendix 7}
\label{S4_Appendix}
\paragraph{Association between alteration detection and credibility}
Across all three studies, we tested whether participants who believed that the video had been digitally altered rated its credibility lower. In each study, credibility differed significantly across the three alteration-detection categories (\emph{Yes}, \emph{No}, \emph{I don't know}) as indicated by Kruskal--Wallis tests (Study~1: $\chi^{2}(2) = 20.77$, $p < .001$; Study~2: $\chi^{2}(2) = 25.28$, $p < .001$; Study~3: $\chi^{2}(2) = 58.96$, $p < .001$). Dunn post-hoc tests consistently showed that participants selecting \emph{Yes} rated the video as less credible than those selecting \emph{No} or \emph{I don't know} (all adjusted $p < .001$ across studies). Collapsing responses into ``Detected'' versus ``Not detected'' confirmed these differences via Wilcoxon rank-sum tests (Study~1: $W = 9778$, $p < .001$; Study~2: $W = 9750.5$, $p < .001$; Study~3: $W = 7999$, $p < .001$). Spearman rank correlations likewise indicated consistent negative associations between alteration detection and credibility (Study~1: $\rho = -0.25$, $p < .001$; Study~2: $\rho = -0.27$, $p < .001$; Study~3: $\rho = -0.40$, $p < .001$). Together, these analyses show a robust pattern across all studies: believing that the video was digitally altered is reliably associated with lower perceived credibility.

\end{document}